\documentclass[fleqn,usenatbib]{mnras}

\usepackage[T1]{fontenc}
\usepackage{ae,aecompl}

\usepackage{graphicx}	
\usepackage{amsmath}	
\usepackage{amssymb}	
\usepackage{wasysym}
\usepackage{hyperref}

\title[ETG density profile and kinematic structure]{Early-type galaxy density profiles from IllustrisTNG: \\ III. Effects on outer kinematic structure}

\author[Y. Wang et al.]{Yunchong Wang$^{1,2,3}$\thanks{E-mail: \url{ycwang19@stanford.edu}},
Shude Mao$^{1}$,
Mark Vogelsberger$^{4}$,
Volker Springel$^{5}$,
Lars Hernquist$^{6}$
\newauthor
and Risa H. Wechsler$^{2,3,7}$
\\
$^{1}$Department of Astronomy, Tsinghua University, Beijing, 100084, China\\
$^{2}$Department of Physics, Stanford University, 382 Via Pueblo Mall, Stanford, CA 94305, USA\\
$^{3}$Kavli Institute for Particle Astrophysics \& Cosmology, P. O. Box 2450, Stanford University, Stanford, CA 94305, USA\\
$^{4}$Kavli Institute for Astrophysics and Space Research, Department of Physics, MIT,  Cambridge, MA 02139, USA\\
$^{5}$Max-Planck-Institut f\"{u}r Astrophysik, Karl-Schwarzschild-Str. 1, D-85748, Garching, Germany\\
$^{6}$Institute for Theory and Computation, Harvard-Smithsonian Center for Astrophysics, 60 Garden Street, Cambridge, MA 02138\\
$^{7}$SLAC National Accelerator Laboratory, Menlo Park, CA 94025, USA
}

\date{Accepted 2022 May 13. Received 2022 May 2; in original form 2021 November 17}

\pubyear{2021}

\begin{document}
\label{firstpage}
\pagerange{\pageref{firstpage}--\pageref{lastpage}}
\maketitle

\begin{abstract}
Early-type galaxies (ETGs) possess total density profiles close to isothermal, which can lead to non-Gaussian line-of-sight velocity dispersion (LOSVD) under anisotropic stellar orbits. However, recent observations of local ETGs in the MASSIVE Survey reveal outer kinematic structures at $1.5 R_{\mathrm{eff}}$ (effective radius) that are inconsistent with fixed isothermal density profiles; the authors proposed varying density profiles as an explanation. We aim to verify this conjecture and understand the influence of stellar assembly on these kinematic features through mock ETGs in IllustrisTNG. We create mock Integral-Field-Unit observations to extract projected stellar kinematic features for 207 ETGs with stellar mass $M_{\ast}\geqslant 10^{11} \mathrm{M_{\astrosun}}$ in TNG100-1. The mock observations reproduce the key outer ($1.5R_{\mathrm{eff}}$) kinematic structures in the MASSIVE ETGs, including the puzzling positive correlation between velocity dispersion profile outer slope $\gamma_{\mathrm{outer}}$ and the kurtosis $h_{4}$'s gradient. We find that $h_{4}$ is uncorrelated with stellar orbital anisotropy beyond $R_{\mathrm{eff}}$; instead we find that the variations in $\gamma_{\mathrm{outer}}$ and outer $h_{4}$ (a good proxy for $h_{4}$ gradient) are both driven by variations of the density profile at the outskirts across different ETGs. These findings corroborate the proposed conjecture and rule out velocity anisotropy as the origin of non-Gaussian outer kinematic structure in ETGs. We also find that the outer kurtosis and anisotropy correlate with different stellar assembly components, with the former related to minor mergers or flyby interactions while the latter is mainly driven by major mergers, suggesting distinct stellar assembly origins that decorrelates the two quantities. 
\end{abstract}




\section{Introduction}
\label{sec:1_intro}

Early-type galaxies~(ETGs, e.g., \citealt{1980ApJ...236..351D, 1987ApJ...313...59D}) are recognized as the `red and dead' end products of hierarchical galaxy formation~\citep{2000MNRAS.319..168C,2001MNRAS.328..726S,2006MNRAS.366..499D,2007MNRAS.375....2D}. Numerical simulations over the past decade have shown that the formation path of early-type galaxies is well-represented by a two-phase scenario~\citep{2007ApJ...658..710N,2008MNRAS.384....2G,2010ApJ...725.2312O,2012ApJ...754..115J,2013MNRAS.428.3121M,2016MNRAS.458.2371R}, comprising an active phase ($z\gtrsim 2$) dominated by gas-rich mergers and bursty in-situ star formation~\citep{2006ApJS..163...50H,2008ApJS..175..356H,2008ApJS..175..390H,2009MNRAS.397..802H,2009ApJ...691.1168H,2015MNRAS.449..361W}, as well as a passive phase ($z\lesssim 2$) dominated by dry mergers and accretion of ex-situ-formed stars~\citep{2009ApJ...703.1531N,2009ApJ...706L..86N,2013ApJ...766...71R,2016MNRAS.456.1030W}, as well as quenching by Active Galactic Nuclei (AGN) feedback~\citep{1998A&A...331L...1S,2003ApJ...596L..27K,2003ApJ...595..614W,2005Natur.433..604D,2005MNRAS.361..776S,2012ARA&A..50..455F,2013ARA&A..51..511K}.

An important feature of ETGs found in observations through strong and weak gravitational lensing~\citep{2006ApJ...649..599K,2007ApJ...667..176G,2009ApJ...703L..51K,2009MNRAS.399...21B,2011MNRAS.415.2215B,2010ApJ...724..511A,2011ApJ...727...96R,2013ApJ...777...98S,2018MNRAS.480..431L,2018MNRAS.475.2403L}, dynamical modeling~\citep{2014MNRAS.445..115T,2015ApJ...804L..21C,2016MNRAS.460.1382S,2017MNRAS.467.1397P,2018MNRAS.476.4543B,2019MNRAS.490.2124L} and X-ray observations of gas dynamics~\citep{2006ApJ...646..899H,2010MNRAS.403.2143H}, is that their total radial matter density profile is well described by a single power-law model $\rho(r)\propto r^{-\gamma^{\prime}}$ with small intrinsic scatter around the slope of $\gamma^{\prime}=2$. This feature is known as the `bulge--halo conspiracy', as neither stars nor dark matter follow a single power-law model with the slope of $\gamma^{\prime}=2$, but their combined profile `conspires' to take the form of a Singular Isothermal Sphere (SIS) (an ideal gas sphere under gravitational-hydrostatic equilibrium):
\begin{equation}
\label{eq:1}
    \rho(r) = \frac{\sigma^2}{2\pi G r^2} \Longleftrightarrow \rho(r)\propto r^{-\gamma^{\prime}},\gamma^{\prime}=2
\end{equation}
where $\sigma$ is the one-dimensional velocity dispersion of stars. In the case of isotropic stellar orbits, $\sigma = \sigma_{r} = \sigma_{\theta} = \sigma_{\phi}$ where the subscripts denote the $(r,\theta, \phi)$ directions in spherical coordinates. In this case, a constant $\sigma$ leads to a constant {\it circular velocity} $v_{c}$ with radius:
\begin{equation}
\label{eq:2}
    v_{c}^2(r) = \frac{G M(<r)}{r} = \frac{G}{r}\int_{0}^{r} \rho(r^{\prime})4\pi r^{\prime 2} dr^{\prime} = 2\sigma^2 .
\end{equation}
Therefore, the ETG total density profile characterized by $\gamma^{\prime}=2$ is naturally linked to its kinematic structure characterized by a flat velocity dispersion/circular velocity radial profile under stellar orbit isotropy. Any deviations from a flat velocity dispersion curve (or the underlying circular velocity) will indicate a deviation of the total density profile from an SIS model ($\gamma^{\prime} \neq 2$). If the 3D velocity dispersion profile can be approximated by a power law $\sigma(r) \propto r^{\alpha}$, then $\gamma^{\prime} < 2$ when $\alpha > 0$ and $\gamma^{\prime} > 2$ when $\alpha < 0$.


However, in the presence of a radial or tangential stellar orbital anisotropy, the {\it projected} stellar velocity dispersion can also vary with radius even when $\gamma^{\prime} = 2$ and $v_{c}$ remains constant with radius. As the anisotropy cannot be measured directly in observations, neither the 3D velocity dispersion profile~\citep{1982MNRAS.200..361B} nor the density profile logarithmic slope~\citep{2008MNRAS.390...71C,2017MNRAS.469.1824X} can be determined in an unbiased manner without an assumed velocity anisotropy. This causes a degeneracy in the derived mass profile/velocity dispersion profile and the assumed velocity anisotropy. Nonetheless, if higher order velocity moments are measured, which provides non-Gaussian information (especially the kurtosis $h_4$, which is the $4^{\mathrm{th}}$ velocity moment) of the LOSVD, this degeneracy can in theory be broken through the opposite behavior of $h_{4}$ under radial and tangential velocity anisotropy~\citep{1992ApJ...391..531D,1993MNRAS.265..213G,1993ApJ...409...75M,2017MNRAS.471.4541R}.

Recent results from the MASSIVE Survey~\citep{2014ApJ...795..158M} found that massive local ETGs with stellar mass $\gtrsim 10^{11.5}\mathrm{M_{\astrosun}}$ that have rising velocity dispersion profiles towards their outskirts tend to have positive $h_4$ and positive $h_4$ gradient~\citep{2017MNRAS.464..356V,2018MNRAS.473.5446V}. \citet{2019ApJ...878...57E} also found that most of MASSIVE ETGs have dropping inner velocity dispersion profiles and increasing $h_4$ towards the galactic center. These trends are in contradiction to the theoretical expectations that under a fixed total density profile, radial velocity anisotropy induces a more positive $h_{4}$  accompanied by a decreasing LOSVD towards the outskirts of a galaxy (vice versa towards the center). To explain this observed tension, \citet{2018MNRAS.473.5446V} proposed that the presence of circular velocity gradients could be the cause, and galaxy-to-galaxy variations of the total density profile slope ($\gamma^{\prime}$) could be present in ETGs. Therefore, we aim to verify this conjecture using simulated ETGs from the state-of-the-art cosmological hydrodynamic simulation IllustrisTNG \citep{2018MNRAS.480.5113M,2018MNRAS.477.1206N,2018MNRAS.475..624N,2018MNRAS.475..648P,2018MNRAS.475..676S}, which possesses a well-studied ETG sample with near-isothermal total density profiles that are broadly consistent with their observed counterparts~(Paper I, \citealt{2020MNRAS.491.5188W}). 

If the outer kinematic structures of ETGs are indeed influenced by the variations in their density profiles, we expect to see a correlation of their outer kinematic structures with minor mergers. This is due to the fact that the evolution of the ETG total density profile at $z<0.5$ is mainly driven by minor mergers as found in \citet{2019MNRAS.490.5722W} (Paper II, hereafter) and similarly in earlier works~\citep{2012ApJ...754..115J,2013MNRAS.429.2924H,2013ApJ...766...71R,2014ApJ...786...89S}. Interestingly, \citet{2019ApJ...874...66G} found strong correlations between $h_4$ and stellar populations probes (i.e. metallicity, metallicity gradients) and suggested cumulative minor mergers might have led to the old-aged, radially-anisotropic ETGs having positive $h_{4}$ at the outskirts. Since galaxy mergers also affect the velocity dispersion profile~\citep{1992ApJ...399..462B,2014ApJ...783L..32S,2020MNRAS.499..559N} and tend to induce radial velocity anisotropy~\citep{2003Sci...301.1696R,2012MNRAS.425.3119H}, we will
also investigate the role of minor mergers in the co-evolution of ETG outer kinematic structure and their density profile using the merger histories of the simulated ETGs. 

This paper is organized as follows: in Section~\ref{sec:2} we introduce the simulation and selection criteria through which we select our ETG catalog, as well as the methods to mimic observations for extracting kinematic structure information out of the ETGs; in Section~\ref{sec:3} we present the results for the kinematic properties of our selected ETG sample along with comparisons to observations; in Section~\ref{sec:4} we further explore the physical interpretation for the formation of outer kinematic structure in ETGs relating to their total density profiles, stellar assembly histories, and environment; in Section~\ref{sec:5} we summarize our main conclusions and provide an outlook for future directions of work. In the following analysis, we assume the Planck-2016 flat-$\Lambda$CDM cosmology~\citep{2016A&A...594A..13P} parameters used by the IllustrisTNG simulations: $h = 0.6774$, $\Omega_{\mathrm{m}} = 0.3089$, $\Omega_{\mathrm{b}} = 0.0486$, $\Omega_{\mathrm{\Lambda}} = 0.6911$, and $\sigma_{\mathrm{8}} = 0.8159$.

\section{Methodology}
\label{sec:2}
\subsection{IllustrisTNG Simulations}
\label{sec:2.1} 

Cosmological simulations have tremendously improved our understanding of galaxy formation and cosmology over the past few decades (see \citealt{2020NatRP...2...42V} for a recent review). {\it The Next Generation} Illustris Simulations\footnote{\url{www.tng-project.org}}~\citep{2018MNRAS.480.5113M,2018MNRAS.477.1206N,2018MNRAS.475..624N,2018MNRAS.475..648P,2018MNRAS.475..676S}, IllustrisTNG for short, is a recent set of cosmo-magneto hydrodynamic simulations evolved with the state-of-the-art moving mesh hydrodynamics code \textsc{Arepo}~\citep{2010MNRAS.401..791S}. They advance the merits of the Illustris Simulations~\citep{2013MNRAS.436.3031V,2014MNRAS.438.1985T}, and improve the Illustris models~\citep{2014MNRAS.444.1518V, 2014Natur.509..177V, 2014MNRAS.445..175G, 2015MNRAS.452..575S,2015A&C....13...12N} in terms of AGN and stellar feedback physics~\citep{2017MNRAS.465.3291W,2018MNRAS.473.4077P}.

The full physics IllustrisTNG simulation suite reproduces many key relations in observed galaxies, including the galaxy-color bimodality in the Sloan Digital Sky Survey~\citep{2018MNRAS.475..624N}, the fraction of dark matter within galaxies at $z=0$~\citep{2018MNRAS.481.1950L}, the galaxy mass-metallicity relation~\citep{2018MNRAS.477L..16T,2019MNRAS.484.5587T} and the intra-cluster metal distribution~\citep{2018MNRAS.474.2073V}, the galaxy size-mass relation evolution~\citep{2018MNRAS.474.3976G}, galaxy morphology transition~\citep{2019MNRAS.487.5416T} and stellar orbital fraction~\citep{2019MNRAS.489..842X}, early-type galaxy total density profiles~\citep{2020MNRAS.491.5188W}, molecular and atomic hydrogen content in low redshift galaxies~\citep{2019MNRAS.487.1529D,2019MNRAS.483.5334S}, star formation activities and quenched fractions~\citep{2019MNRAS.485.4817D}, ram-pressure stripping in dense environments~\citep{2019MNRAS.483.1042Y}, gas-phase
metallicity gradients in star-forming galaxies ~\citep{2021MNRAS.506.3024H}, as well as AGN galaxy occupation and X-ray luminosities~\citep{2018MNRAS.479.4056W,2019MNRAS.484.4413H,2020MNRAS.493.1888T}. Although some facets of these comparisons still exhibit discrepancies with observations to different levels, the significant improvements over Illustris and the multitude of agreement in galaxy and cluster level properties with observations demonstrates the predictive power of IllustrisTNG (e.g., predictions of JWST observation for high redshift galaxies~\citealt{2020MNRAS.492.5167V}). Therefore, we use ETGs selected from IllustrisTNG to gain insights on the origin of their outer kinematic structure as seen in the MASSIVE Survey.


The simulation suite of IllustrisTNG comprises 3 cubic boxes with periodic boundary conditions, i.e. TNG100 (side length $75\,\mathrm{Mpc}/h$, same as original Illustris), TNG300 (side length $205\,\mathrm{Mpc}/h$), and TNG50 (side length $35\,\mathrm{Mpc}/h$), with overall higher resolution in smaller boxes and each box contains several runs with different numerical resolutions. In this paper, we select galaxies from the highest resolution run of the TNG100 box, which is best for our purpose of studying ETGs with stellar mass $\gtrsim 10^{11}\mathrm{M_{\astrosun}}$, since it provides a substantial sample size of ETGs in this mass range with reasonable mass resolution. TNG100 has a baryonic matter mass resolution of $m_{\mathrm{baryon}} = 1.4\times10^{6} \mathrm{M}_{\mathrm{\astrosun}}$ and a dark matter mass resolution of $m_{\mathrm{DM}} = 8.9\times 10^{6} \mathrm{M}_{\mathrm{\astrosun}}$, each with $1820^{3}$ resolution elements. The softening length scale of dark matter and stellar particles is $\epsilon = 0.74\,\mathrm{kpc}$ (valid for $z \leq 1$, scales as $(1+z)^{-1}\epsilon$ at $z>1$), whereas the gravitational softening of the gas cells is adaptive and has a minimum length scale of 0.19 comoving kpc. The simulation runs of TNG50, TNG100 and TNG300 boxes are now available for public data access~\citep{2019ComAC...6....2N}. 

\subsection{Sample selection}
\label{sec:2.2}

Galaxies in IllustrisTNG are identified as gravitationally-bound structures (subhalos) by \textsc{Subfind}~\citep{2001MNRAS.328..726S,2009MNRAS.399..497D} wiin halos found using the Friends of Friends (FoF) algorithm based on mean particle separation length. The largest subhalo in a halo together with its baryonic component is defined as the central galaxy, and all other subhalos in the halo are defined as satellite galaxies. We select central galaxies in TNG100 with total stellar masses of $M_{\ast}\geqslant 10^{11}\mathrm{M_{\astrosun}}$ ($\gtrsim 7\times10^4$ resolution elements), which covers the stellar mass range of the MASSIVE ETGs ($M_{\ast}\geq 3\times 10^{11}\mathrm{M_{\astrosun}}$, \citealt{2018MNRAS.473.5446V}). Merger trees that trace the comprehensive assembly histories of galaxies and dark matter halos are constructed using the algorithm \textsc{Sublink}~\citep{2015MNRAS.449...49R}. 

\begin{figure*}
	\includegraphics[width=2.08\columnwidth]{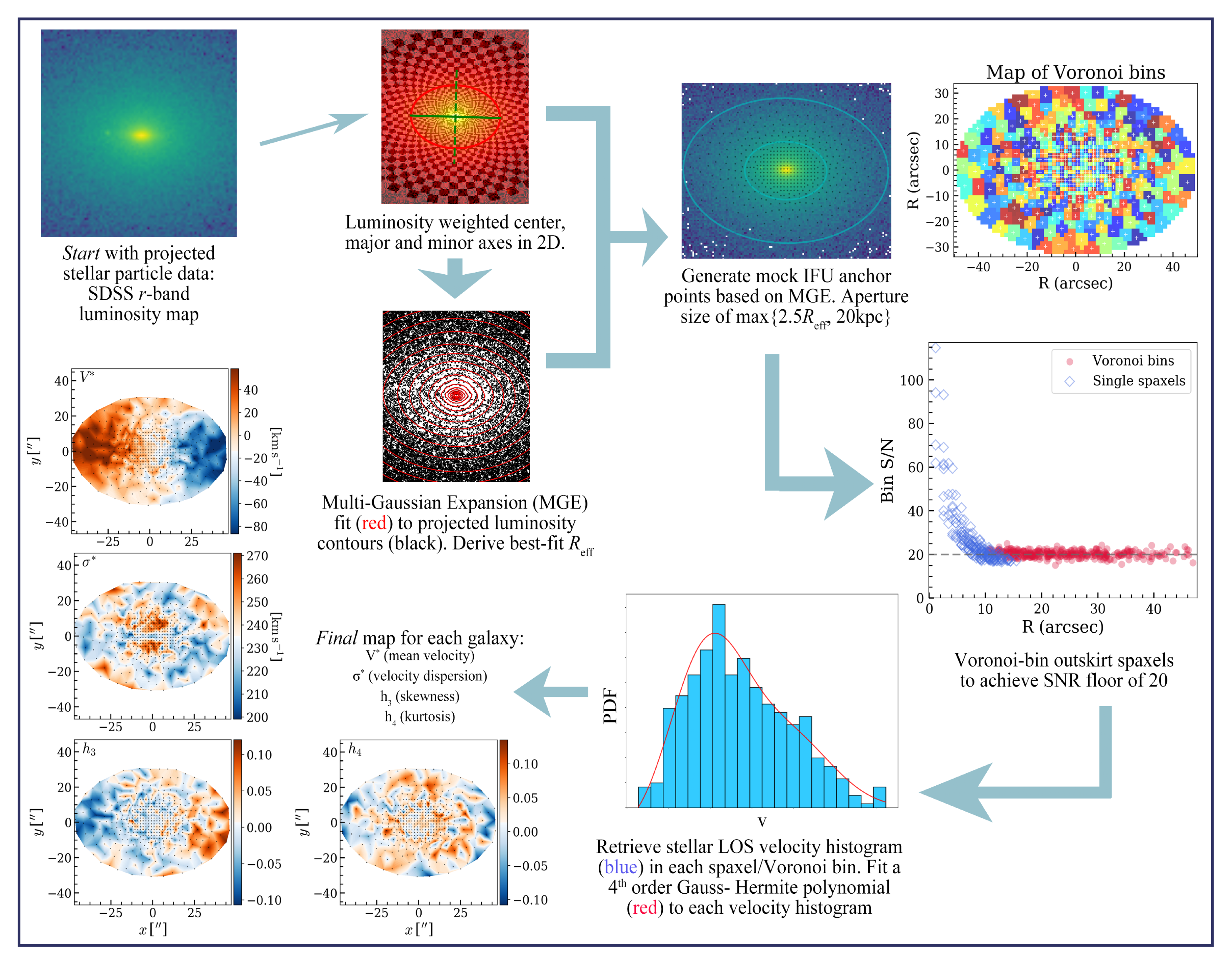}
    \caption{Schematic flow chart of the main steps for our mock-IFU post-processing pipeline. The input starts from the projected stellar luminosity map from particle data, and the output is the 2D kinematic structure maps for each galaxy, i.e. mock IFU data products. The example galaxy (\textsc{Subfind} ID = 257302) shown has typical elliptical geometry in its luminosity map. The central projected region is well resolved with a large number of high SNR spaxels, and there are increasing fractions of Voronoi bins towards the outskirts. The velocity histogram indicates one of the spaxel/Voronoi bin demonstrating a  significant non-Gaussian LOS velocity profile. Comparing the mean velocity and dispersion maps shows that the galaxy is dominated by randomized stellar motion, while its skewness $h_{3}$ distribution is anti-correlated with the mean velocity distribution, and kurtosis $h_{4}$ is generally positive for most of the spaxels/Voronoi bins in the map.}
    \label{fig:1}
\end{figure*}

We follow Paper I and Paper II for galaxy morphology classification, which is based on the optical luminosity reconstruction approach detailed in \citet{2017MNRAS.469.1824X}. The optical light of a galaxy is derived using the Stellar Population Synthesis (SPS) model~\citep{2003MNRAS.344.1000B} based on the metallicity and age of its stellar particles which are treated  as coeval stellar populations with Chabrier initial mass function~\citep{2003PASP..115..763C}. A projection dependent dust attenuation is then applied to the galaxy luminosity and we take the SDSS $r$-band luminosity after dust processing to calculate azimuthally-averaged galaxy luminosity radial profiles for morphological classification. We fit single-component models, i.e. the de Vaucouleurs profile (S$\mathrm{\acute{e}}$rsic $n=4$) and the exponential profile (S$\mathrm{\acute{e}}$rsic $n=1$), as well as a combined two-component model of exponential and de Vaucouleurs profiles to the projected radial luminosity profiles of galaxies. If a galaxy is better-fit (lower minimum $\chi^2$) by the de Vaucouleurs profile, and demonstrates bulge-dominance in the two-component fit (bulge-to-total ratio $>0.5$) in all three independent projections (along the $x$, $y$, and $z$ axes of the simulation box), then it is considered as early-type. 

To make our ETG classification more robust according to Integral-Field-Unit (IFU hereafter) observations~\citep{2018MNRAS.476.1765L}, we fit a single S$\mathrm{\acute{e}}$rsic profile to the projected luminosity profiles of the selected ETGs, and keep only those with S$\mathrm{\acute{e}}$rsic indices satisfying $n_{\mathrm{x}}\geq 2.5$
, $n_{\mathrm{y}}\geq2.5$, and $n_{\mathrm{z}}\geq 2.5$ in all three projections simultaneously (\citealt{2018MNRAS.476.1765L} selected ETGs based on $n\geq 2.5$). For the MASSIVE ETGs~\citep{2014ApJ...795..158M}, they were selected based on morphology (E and S0 types from \citealt{2003A&A...412...45P}) without any specific S$\mathrm{\acute{e}}$rsic index cut applied, and 77 out of 105 galaxies have $n$ ranging from 2 to 6 cross referencing the NSA catalog. Therefore, we consider our generic ETG selection criteria a reasonable choice when comparing to MASSIVE ETGs and we arrive at a sample of 221 well-resolved ETGs after the above-mentioned photometric selections. This will be further reduced to a {\it final} sample of 207 ETGs after removing galaxies that have been through recent major mergers  (see Section~\ref{sec:3.1}).
 
\subsection{Mock observations}
\label{sec:2.3}

We summarize the main steps in post-processing simulation particle and catalog data to retrieve kinematic properties of our mock ETG sample that mimics kinematic properties from the observational Integral-Field-Unit (IFU) spectroscopic surveys. Our pipeline is largely based upon the public code $\textsc{illustris-tools}$~\footnote{\url{https://github.com/HongyuLi2016/illustris-tools}}~\citep{2016MNRAS.455.3680L} to make mock IFU observations, with edits for our IllustrisTNG ETG sample applied. Fig.~\ref{fig:1} shows the work flow of our post-processing pipeline. 

We find the center of the stellar component of the galaxy by comparing the center of mass for all stellar particles in the galaxy and its central 20$\%$ of particles (in 3D radial distance). We do this recursively by setting the central 20$\%$ as the total region considered to be the total region in the next step, and compare its center of mass to the 20$\%^2$ particles center of mass, until the difference between the two center of masses in a step drops below 0.01 kpc. This center for the stellar component is within $\sim 0.05$ kpc of the minimum gravitational potential point defined by $\textsc{Subfind}$ for our ETG sample. We assume the mean velocity of the stellar particles to be the mass-weighted average of particle velocities within 15 kpc of the stellar component center found as above. We calibrate coordinates and velocities of stellar particles with respect to the stellar component center and mean velocity of their galaxy. 

Afterwards, we make 2D projections of the stellar particles in random directions (along the z axis of the simulation box) to mimic realistic observational conditions (we have also tried random projections in the x and y axis, as well as along the edge-on direction of the galaxy, which all largely preserve the main findings of this paper). We pixelize the SDSS $r$-band luminosities of the projected stellar particles onto an $80\,\mathrm{kpc}\times 80\,\mathrm{kpc}$ square aperture centered on the stellar component center. We set our pixel size to $0.5\,\mathrm{kpc}\times 0.5\,\mathrm{kpc}$, imitating the local ETG population of the MASSIVE survey, and we place our fiducial ETGs at $z=0.03$ (angular diameter distance 128 Mpc) for our mock observations. The angular resolution of our mock luminosity map is $0.806^{\prime\prime}$ at that redshift, while the MASSIVE galaxies were observed using the Hobby-Eberly Telescope~(HET, \citealt{HET2008}) at the McDonald Observatory, where its operational seeing has a FWHM of $1.5^{\prime\prime}$ ($\sigma_{\mathrm{PSF}} = 0.64^{\prime\prime}$)
If we convolve a Gaussian PSF kernel matching the HET seeing, which has a Gaussian scatter than our mock pixel size, the PSF normalization will boost each pixel's value by itself and does not smear out its flux to other neighbouring pixels. As such, we consider our pixelization has effectively accounted for the seeing with it being marginally coarser than the expected PSF. 

Next, we model the projected luminosity map with the Multi-Gaussian-Expansion (MGE) formalism~\citep{1994A&A...285..739E} with the publicly available python package $\textsc{MgeFit}$\footnote{\url{https://www-astro.physics.ox.ac.uk/~mxc/software/}}~\citep{2002MNRAS.333..400C}. The MGE method models the surface brightness of a galaxy with a stack of (we choose $N=15$) elliptical 2D Gaussians:
\begin{equation}
\label{eq:3}
	\Sigma(x^{\prime}, y^{\prime}) = \sum_{k = 1}^{N}\frac{L_{k}}{2\pi\sigma_{k}^{2}q_{k}^{\prime}}\mathrm{exp}\left[-\frac{1}{2\sigma_{k}^{2}}\left(x^{\prime2} + \frac{y^{\prime2}}{q_{k}^{\prime2}}\right)\right],
\end{equation}
where $L_k$, $\sigma_k$, $q_{k}^{\prime}$ are the normalization, standard deviation, and axis ratio of the $k$-th Gaussian. The primed coordinates $x^{\prime}$, $y^{\prime}$ are the projected 2D angular coordinates (placing the galaxy at 128 Mpc) in a system where the origin is at the center of the galaxy and $x^{\prime}$ axis being aligned with the galaxy's major axis~\citep{2002MNRAS.333..400C}. The center and major (minor) axis of the galaxy's projected luminosity map are calculated using the top $10\%$-brightest pixels~\citep{2016MNRAS.455.3680L}, with the center being determined by luminosity-weighting and the major (minor) axis following the eigenvectors of the 2D inertial tensor~\citep{2006MNRAS.367.1781A} spanned by the selected pixels.

The luminosity profile is then sampled in equal angular bins and logarithmic radial bins to perform the MGE fit. The result of this step produces an analytical best-fit description of the galaxy surface brightness map with 2D Gaussians, and also provides realistic effective radius (half-light radius) of our mock ETGs that mimics observations by integrating the best-fit surface brightness profiles. A comparison between the MGE-derived effective radius $R_{\mathrm{eff}}^{\mathrm{MGE}}$ and the effective radius obtained by directly projecting all stellar particles assigned to the galaxy by $\textsc{Subfind}$ ($R_{\mathrm{eff}}$) is shown in Appendix~\ref{sec:App_A}, which shows that $R_{\mathrm{eff}}^{\mathrm{MGE}}$ gives a more realistic description of galaxy sizes removing much of the intra-cluster light especially in more massive ETGs. This choice of effective radius is also more flexible to capture different distinct components in the ETG luminosity profile compared to simpler half-light-ellipse or single S$\mathrm{\acute{e}}$rsic fits often adopted in observations (see Section 3.2 in \citealt{2014ApJ...795..158M}). In the rest of this paper, we refer to $R_{\mathrm{eff}}^{\mathrm{MGE}}$ when we mention the effective radius of our simulated ETGs. 

\begin{figure*}
	\includegraphics[width=2\columnwidth]{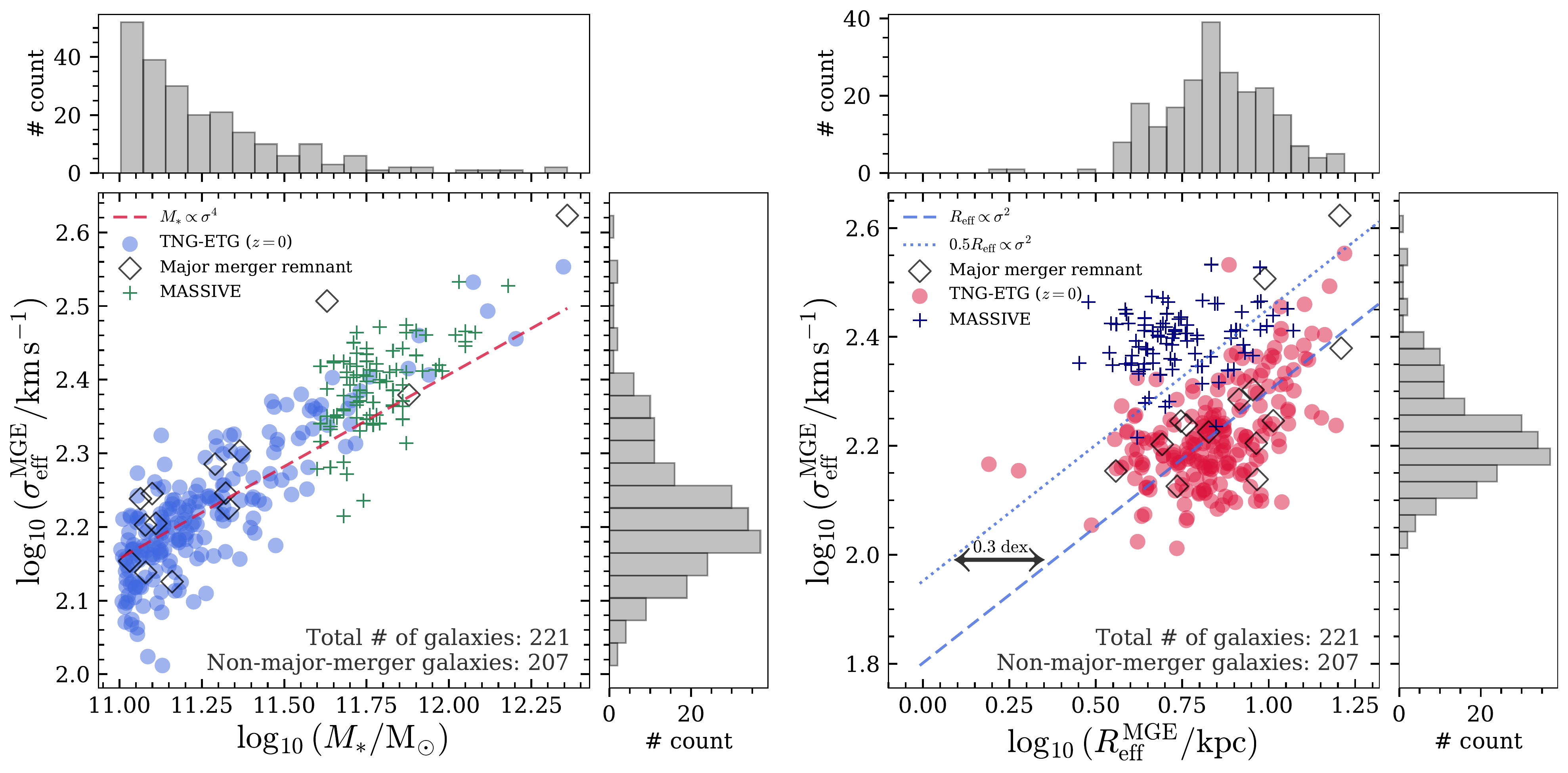}
    \caption{The distribution of stellar mass (left panel) and effective radius (right panel) versus the velocity dispersion measured within the effective radius. The open markers indicate those ETGs that are either undergoing major mergers found via their kinematic maps, or that have been through a recent one (mass ratio $>1/3$) in the previous snapshot recorded in the merger tree. We remove them in the following analysis and figures following~\citet{2018MNRAS.473.5446V}, and only include the filled marker points in our samples below (207 galaxies remaining out of 221). The histograms at the sides in each subplot show the marginalized distributions of these three quantities (histograms include the major merger galaxies). Dashed lines in both panels indicate scaling relations (fixed slope and fitted intercept) for ETGs as expected from observations. `Plus' markers in both panels indicate the MASSIVE sample in \citet{2018MNRAS.473.5446V}. Our simulated ETG sample has only 18 galaxies with $\log_{10}(M_{\ast}/ \mathrm{M_{\astrosun}}) \geqslant 11.6$) versus 89 in MASSIVE, owing to the volume limit of the simulation. The effective radius for MASSIVE ETGs quoted here are from the 2MASS catalog (which underestimates galaxy sizes by a factor $\sim2\ (0.3\ \mathrm{dex})$ compared to the NSA catalog, but covers the entire MASSIVE sample) as summarized in \citet{2014ApJ...795..158M}. We also show a dotted line in the right panel that indicates a scaling relation with half the effect radius of the fitted IllustrisTNG ETG scaling which coincides well with the MASSIVE galaxies.}
    \label{fig:2}
\end{figure*}

Furthermore, we generate mock IFU maps for our simulated ETGs focused on their central regions satisfying:
\begin{equation}
\label{eq:4}
	r^{\prime} = \sqrt{x^{\prime 2} + \frac{y^{\prime 2}}{q_\mathrm{p}^2}} < \left(\frac{R_{\mathrm{max}}}{\sqrt{q_{\mathrm{p}}}}\right),
\end{equation}
where $R_{\mathrm{max}} = \mathrm{max}\{2.5 R_{\mathrm{eff}}^{\mathrm{MGE}}, 20\,\mathrm{kpc}\}$ is our mock IFU aperture that mimics observations~\citep{2017MNRAS.464..356V}. This aperture setting guarantees that the kinematic maps we generate sample the outskirts of the galaxies well. After we select these central pixels, we assume that each individual pixel corresponds to an IFU spaxel (single fiber in an IFU bundle). We use the $\textsc{Convex-Hull}$ method to efficiently identify stellar particles that are projected within our IFU aperture, and assign them to their nearest spaxel by querying the KD-Tree constructed for all spaxel anchor points. Following that, we Voronoi-bin the outskirt spaxels to the target signal-to-noise ratio (SNR) of 20 to ensure kinematic information quality, and leave single spaxels in the central region with high SNR unbinned. 

Finally, we calculate the line-of-sight (LOS) stellar velocity distribution for each spaxel/Voronoi-bin in the 2D IFU aperture. If a spaxel/Voronoi-bin has $N$ stellar particles projected within it, then we construct the LOS velocity histogram of these particles using $\left[ \sqrt{N} \right]$ equi-velocity bins. We perform a least-squares fit to the stellar LOS velocity histogram in each spaxel/Voronoi-bin with a fourth-order Gauss-Hermite function to extract the mean ($\bar{v}$), dispersion ($\sigma$), skewness ($h_{3}$), and kurtosis ($h_{4}$) of the LOS velocity distribution:
\begin{equation}
\label{eq:5}
\begin{split}
	f(v) &= \frac{1}{\sqrt{2\pi \sigma^{2}}}\,\mathrm{exp}\left(-\frac{x^2}{2}\right) \\
	&\quad\quad \times \left[1 + h_{3}\mathrm{H_{3}}\left(x\right) + h_{4}\mathrm{H_{4}}\left(x\right)\right], x = \frac{v - \bar{v}}{\sigma}, 
\end{split}
\end{equation}
where $\bar{\nu}$ and $\sigma$ are the mean velocity and dispersion in the spaxel/Voronoi bin. $\mathrm{H_{3}}(x)= (2/\sqrt{3})x^{3} - \sqrt{3}x$ and $\mathrm{H_{4}}(x) = (\sqrt{6}/3)x^{4} - \sqrt{6}x^{2} + \sqrt{6}/4$ are the third and fourth order normalized Gauss-Hermite functions, while their coefficients contain the important kinematic information of the third velocity moment $h_3$ (skewness) and the fourth velocity moment $h_4$ (kurotsis). Qualitatively, a positive $h_{3}$ indicates asymmetric bias of the velocity distribution towards velocities less than the mean, and vice versa for negative $h_{3}$. A positive $h_{4}$ indicates symmetric deviation from a normal Gaussian with larger tails and a more centrally peaked velocity distribution, while negative $h_{4}$ indicates smaller tails and centrally flat (sometimes double peaked) velocity distribution~\citep{1993ApJ...407..525V}. To wrap up, we store the 2D map of the  mean, dispersion, skewness, and kurtosis of the line-of-sight (LOS) velocity distribution for each individual ETG, and proceed to obtaining radial profiles of these kinematic properties in the next section.

\section{Results of kinematic features}
\label{sec:3}

\begin{figure*}
	\includegraphics[width=1.9\columnwidth]{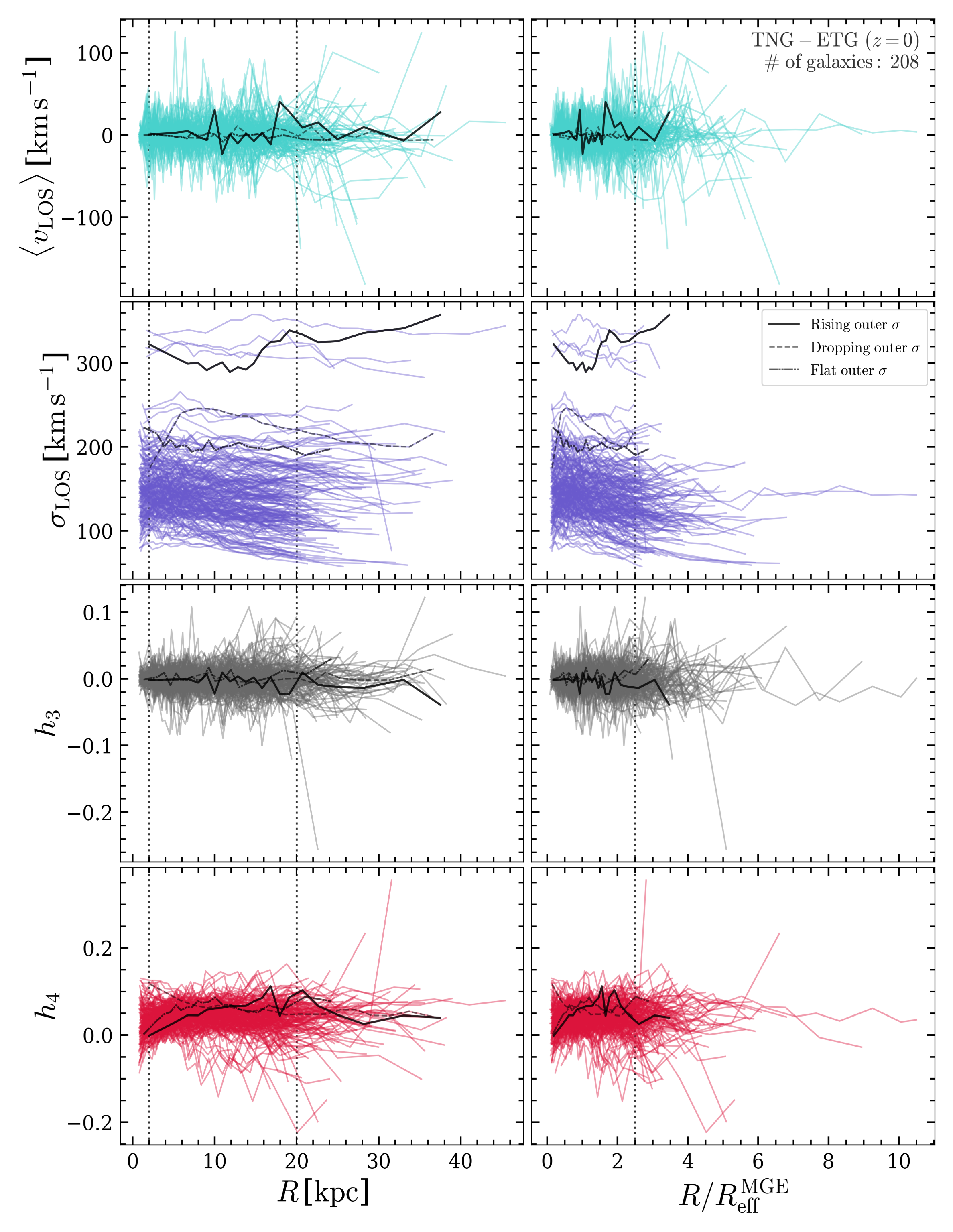}
    \caption{From top to bottom, the profiles of the line of sight mean velocity ($\langle v_{\mathrm{LOS}} \rangle$), velocity dispersion $\sigma_{\mathrm{LOS}}$, skewness ($h_{3}$), and kurtosis ($h_{4}$) are shown for each individual galaxy in our ETG sample. Each curve represents one galaxy's flux-weighted radial profile of the four kinematic quantities. The left row shows the profile with respect to the projected physical distance to the galaxy center, while the right row shows the results w.r.t. the projected distance in units of the effective radius $R_{\mathrm{eff}}^{\mathrm{MGE}}$ for each individual galaxy. The vertical dotted lines in the left column indicate the location of $R = 2$ kpc and $R = 20$ kpc, while the vertical dotted lines in the right column indicate the location of $R = 2.5 R_{\mathrm{eff}}^{\mathrm{MGE}}$. We select three individual galaxies that have rising (solid), flat (dotted-dashed), dropping (dashed) outer velocity dispersion profiles at 20 kpc (see Section~\ref{sec:3.2} for definitions), and highlight their kinematic profiles in each panel.}
    \label{fig:3}
\end{figure*}

\subsection{The galaxy sample}
\label{sec:3.1}

In this section, we present the fundamental properties of our selected ETG sample. Fig~\ref{fig:2} shows the stellar mass ($M_{\ast}$), the effective radius derived with the best-fit MGE model ($R_{\mathrm{eff}}^{\mathrm{MGE}}$), and the mass-weighted average LOS velocity dispersion ($\sigma_{\mathrm{eff}}^{\mathrm{MGE}}$) within $R_{\mathrm{eff}}^{\mathrm{MGE}}$. The sample is a typical massive (median stellar mass $\log_{10}\left(M_{\ast}/\mathrm{M_{\astrosun}}\right)=11.18$) ETG sample with well-resolved spatial extent (median effective radius of 7 kpc), and dynamically hot with significant velocity dispersion (median velocity dispersion $162\,\mathrm{km\,s^{-1}}$). The LOS velocity dispersion within the effective radius positively correlates with stellar mass following the Tully--Fisher relation $M_{\ast} \propto \sigma^{4}$~\citep{2000ApJ...533L..99M,2001ApJ...550..212B,2010MNRAS.409.1330W}. It also correlates with the effective radius, which is observed to roughly follow $R_{\mathrm{eff}} \propto M_{\ast}^{0.5}$ in the local universe~\citep{2013ApJ...775..106C,2013MNRAS.428.1715H,2015MNRAS.447..636X}, and leads to a scaling with the LOS stellar velocity dispersion in the form of $R_{\mathrm{eff}}\propto \sigma^{2}$. We fit these two scaling relations to our IllustrisTNG ETG sample with fixed power-law slopes and free intercepts (shown as dashed lines in Fig.~\ref{fig:2}). We also plot in Fig.~\ref{fig:2} the stellar masses, effective radii, and the average LOS stellar velocity dispersion for MASSIVE galaxies in \citet{2018MNRAS.473.5446V}. The $M_{\ast}$-$\sigma$ scaling of MASSIVE galaxies is rather consistent with the IllustrisTNG ETGs, although IllustrisTNG approaches the volume limit for massive galaxies (only 18 galaxies with $\log_{10}(M_{\ast}/ \mathrm{M_{\astrosun}}) \geqslant 11.6$) versus the MASSIVE Survey (89 galaxies with $\log_{10}(M_{\ast}/ \mathrm{M_{\astrosun}}) \geqslant 11.6$). The effective sizes quoted in the right panel of Fig.~\ref{fig:2} are from the 2MASS catalog which covers essentially all MASSIVE ETGs. However, due to the shallower survey depth of 2MASS, galaxy sizes are underestimated by a factor $\sim2$ due to insufficient sensitivity to galaxy outskirts compared to sizes from the NSA catalog which is based on SDSS DR8 photometry~(see Section 3.2 in \citealt{2014ApJ...795..158M}). We also plot a scaling relation that shifts the IllustrisTNG ETG $R_{\mathrm{eff}}$-$\sigma$ scaling by half the effective radii in Fig.~\ref{fig:2} which coincides well with the MASSIVE ETGs.

Before diving in to the results, we would like to point out that the selection of galaxy samples in the MASSIVE Survey~\citep{2014ApJ...795..158M,2018MNRAS.473.5446V} has deliberately removed galaxies in ongoing mergers or show complex merger remnant structures. We therefore also remove such out-of-equilibrium galaxies from our sample to further make fair comparisons with the observations. First, we remove galaxies at $z=0$ that are currently going through major mergers apparent from their kinematic structure (mainly LOS mean velocity) and luminosity maps. Second, we also trace every ETG in our sample back one snapshot from $z=0$ to $z\approx 0.01$ (140 Myrs ago) along their main progenitor branch, and remove all galaxies that are remnants of recent major mergers in that snapshot with merger stellar mass ratio $\mu_{\ast} \geqslant 1/3$. Combining these two removal criteria, our original 221 ETG sample reduces to 207 galaxies, and the removed galaxies are indicated by empty diamonds in Fig.~\ref{fig:2}. We proceed with our analysis in the following with the final 207 galaxies without significant perturbation.

\begin{figure}
	\includegraphics[width=\columnwidth]{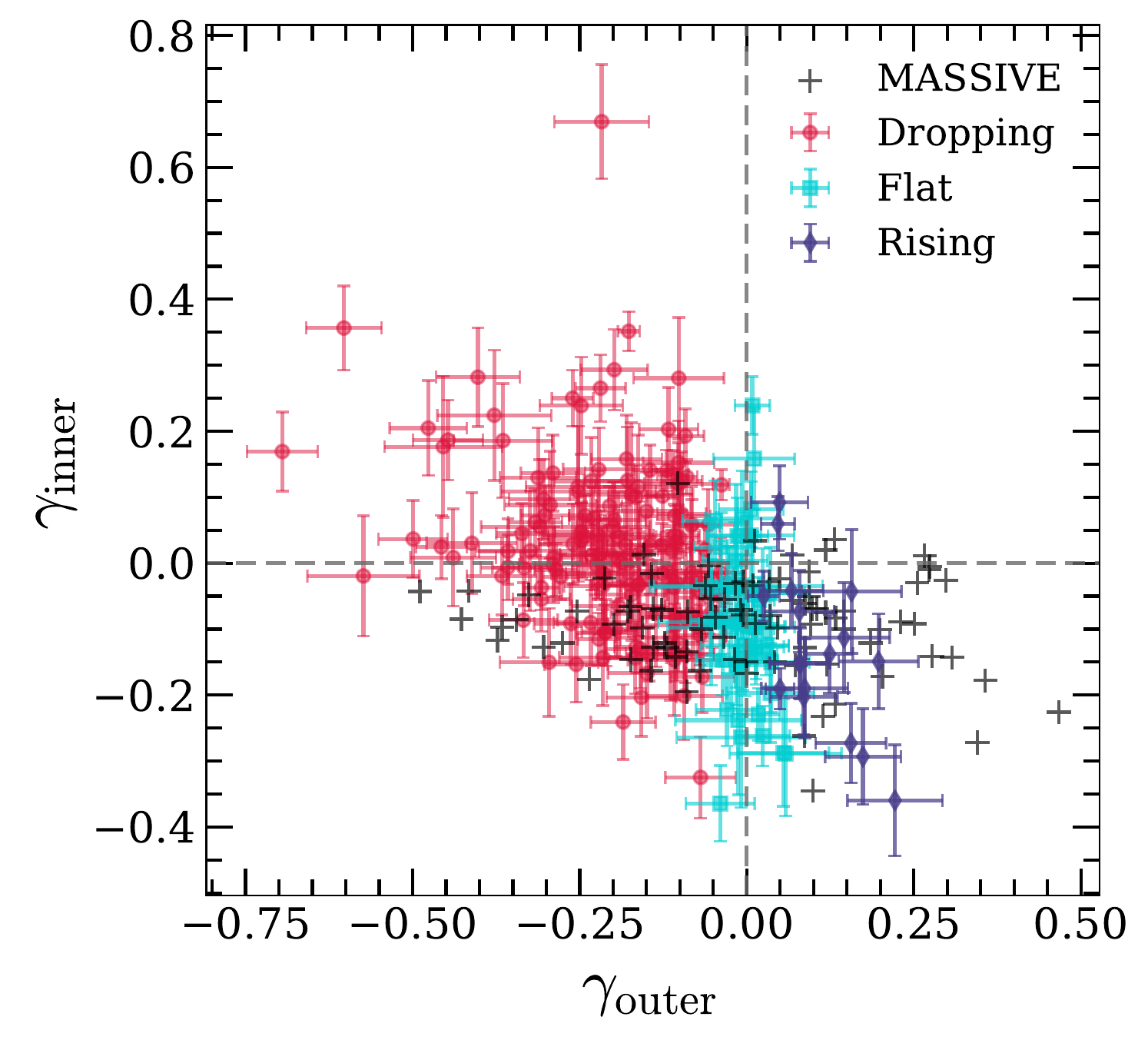}
    \caption{The distribution of the inner and outer slopes of the velocity distribution profile. The point (0, 0) indicates the case for a single-power-law isothermal density profile. The colored points indicate simulated ETGs with rising (blue), flat (green), and dropping (red) outer velocity dispersion profile slopes ($\gamma_{\mathrm{outer}}$). The error bars of the points are the combined fitting errors of the velocity dispersion profile for the inner and outer slopes. Observation values from the MASSIVE Survey~\citep{2018MNRAS.473.5446V} are shown for comparison with plus sign markers.}
    \label{fig:4}
\end{figure}

In Fig.~\ref{fig:3} we show the projected profiles of the mean velocity ($\langle v_{\mathrm{LOS}} \rangle$), the stellar velocity dispersion ($\sigma_{\mathrm{LOS}}$), the skewness ($h_{3}$), and kurtosis ($h_{4}$) of our 207 ETGs sample. We bin the spaxels/Voronoi bins of the projected map radially for each galaxy in 20 equal-number bins, and calculate the flux-weighted (SDSS $r$-band) value of the four quantities in each bin to represent the value for the average velocity profile in that radial bin. The left column has the projected physical distance $R$ as the x axis, while the right column has the scaled projected distance in units of $R_{\mathrm{eff}}^{\mathrm{MGE}}$ as x-axis. We can see that in both radial scalings the mean velocity and the skewness profiles are largely flat, while the LOS velocity dispersion profile takes the form of a variety of shapes, and that the $h_{4}$ profiles show a mildly increasing trend towards larger radii. 

\subsection{Inner and outer slopes of the velocity dispersion profile}
\label{sec:3.2}

To measure the slopes of the projected velocity dispersion profile and compare with the results of~\citet{2018MNRAS.473.5446V}, we follow their modeling by fitting a double-power-law model to the dispersion profile. The dispersion profile model takes the form:
\begin{equation}
    \label{eq:6}
    \sigma_{\mathrm{fit}}(R) = \sigma_{0} 2^{\gamma_{1} - \gamma_{2}} \left(\frac{R}{R_{\mathrm{B}}}\right)^{\gamma_{1}} \left(1+\frac{R}{R_{\mathrm{B}}}\right)^{\gamma_{2} - \gamma_{1}},
\end{equation}
where $\gamma_{1}$ and $\gamma_{2}$ are the asymptotic slopes of the profile towards zero and infinity, $R_{\mathrm{B}}$ the break radius is set to 5 kpc following \citet{2018MNRAS.473.5446V}, and $\sigma_{0}$ is the normalization of the dispersion profile. We have verified that the break radius of 5 kpc is a reasonable description of the IllustrisTNG ETGs. We fit each galaxy's radial velocity dispersion profile in the radial range [$2\,\mathrm{kpc}$, $R_{\mathrm{max}}$] (see Equation~\ref{eq:4} for $R_{\mathrm{max}}$) to avoid core softening in the central regions of the ETGs due to the resolution limit of the simulation. With an analytic double-power-law description of the dispersion profiles, we can define the logarithmic slope of the fitted dispersion profile at any radius $R$ as:
\begin{equation}
    \label{eq:7}
    \gamma(R) = \frac{d \mathrm{log}\, \sigma_{\mathrm{fit}}(R)}{d \mathrm{log} R} = \gamma_{1} + \left(\gamma_{2} - \gamma_{1}\right) \left(\frac{R}{R_{\mathrm{B}}}\right) \left(1 + \frac{R}{R_{\mathrm{B}}}\right)^{-1}.
\end{equation}
Since in the MASSIVE Survey, the inner and outer slopes of the dispersion profile are measured at $R=2$ kpc and $R=20$ kpc, we follow that definition and obtain the inner ($\gamma_{\mathrm{inner}}$) and outer ($\gamma_{\mathrm{outer}}$) slopes of the velocity dispersion profile:
\begin{equation}
    \label{eq:8}
    \gamma_{\mathrm{inner}} = \gamma(2) = \frac{5}{7} \gamma_{1} + \frac{2}{7} \gamma_{2},\ \gamma_{\mathrm{outer}} = \gamma(20) = \frac{1}{5} \gamma_{1} + \frac{4}{5} \gamma_{2}.
\end{equation}
Using the slopes $\gamma_{1}$ and $\gamma_{2}$ obtained from the best double-power-law fit in Equation~\ref{eq:6}, we can derive the  $\gamma_{\mathrm{inner}}$ and $\gamma_{\mathrm{outer}}$ values for our ETGs accordingly. We also follow uncertainties introduced in the fitting procedure to $\gamma_{1}$ and $\gamma_{2}$, and propagate them appropriately to uncertainties in $\gamma_{\mathrm{inner}}$ and $\gamma_{\mathrm{outer}}$. Galaxies consistent with $\gamma_{\mathrm{outer}} = 0$ within the uncertainties are defined having `flat' outer dispersion profiles (0 within upper and lower bounds of $\gamma_{\mathrm{outer}}$), galaxies with lower bound of $\gamma_{\mathrm{outer}} > 0$ are defined as having `rising' outer dispersion profiles, and galaxies with upper bound $\gamma_{\mathrm{outer}} < 0$ are defined as having `dropping' outer dispersion profiles, consistent with \citet{2018MNRAS.473.5446V}.

The distribution of the inner and outer slopes (along with their uncertainties) of the dispersion profiles of our ETG sample are shown in Fig.~\ref{fig:4}. Galaxies with flat, rising, or dropping outer dispersion profiles are indicated with differently-colored markers, and we also over-plot the best-fit values of the observed MASSIVE ETGs from \citet{2018MNRAS.473.5446V}. From the figure we see that the $\gamma_{\mathrm{outer}}$ of the IllustrisTNG ETGs generally agree with the range of outer slopes in the observations. The simulation also has similar levels of uncertainties in the slopes ($\sim 0.1$, not shown for the observed values) as the observations. The range of $\gamma_{\mathrm{inner}}<0$ values for our ETGs is also very similar to observations, however, it is noticeable that there are apparently more galaxies having $\gamma_{\mathrm{inner}} > 0$ within the uncertainties in our sample, which could be produced by systematic underestimation of the central stellar velocity dispersion. Since we mainly focus on the behavior of $\gamma_{\mathrm{outer}}$ in this work, the discrepancy in $\gamma_{\mathrm{inner}}$ with observations does not affect our findings below. Nevertheless, this systematic underestimation of the central stellar velocity dispersion again points to the fact that IllustrisTNG galaxies are subject to higher central dark matter fractions~\citep{2018MNRAS.481.1950L} and halo contraction~\citep{2020MNRAS.491.5188W}, which adds another galaxy property that needs to be anchored by future improvements of the underlying feedback physics model of the simulation.

\subsection{Higher order velocity moments $h_{3}$ and $h_{4}$}
\label{sec:3.3}

\begin{figure*}
	\includegraphics[width=2\columnwidth]{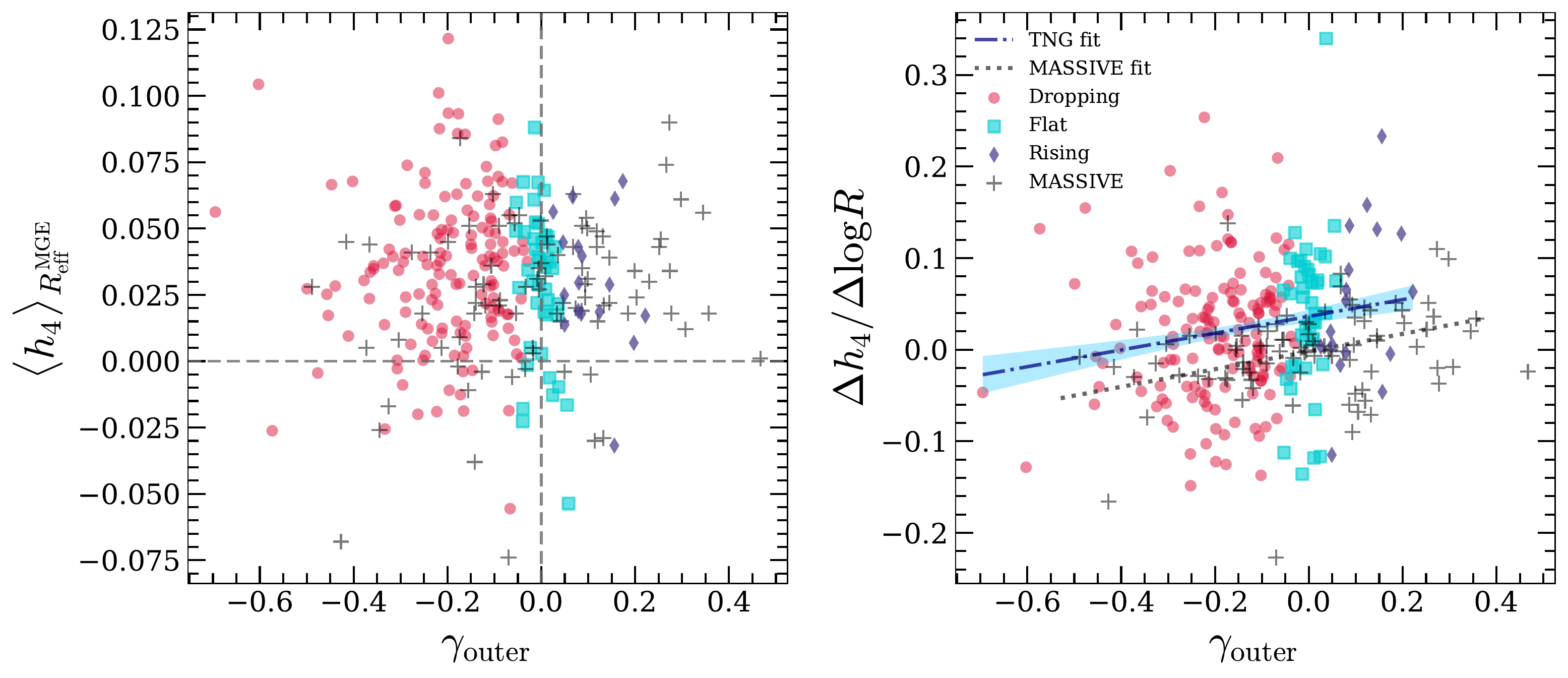}
    \caption{The average kurtosis (left panel, measured within the effective radius) and kurtosis gradient (right panel, measured within $0.5R_{\mathrm{eff}}^{\mathrm{MGE}} < R < 1.5R_{\mathrm{eff}}^{\mathrm{MGE}}$) versus the velocity dispersion profile outer slope $\gamma_{\mathrm{outer}}$. The colored dots indicate the three types of simulated ETGs with different outer slopes of their velocity dispersion profiles, while the open markers are observations from the MASSIVE Survey. The point (0, 0) in the left panel indicates the point where the line-of-sight velocity distribution is a perfect Gaussian and independent of the projected radius. In the right panel, the dotted line indicates the best linear fit for the MASSIVE ETGs (from Fig. 6 in \citealt{2018MNRAS.473.5446V}, slope $b = 0.096\pm0.037$, $p=0.012$), and the blue dotted-dashed line indicates the best linear fit for the IllustrisTNG ETGs (slope $b = 0.092\pm0.035$, $p=0.009$, shaded region stands for $68\%$ confidence interval). Clearly, the IllustrisTNG ETGs also show positive correlation between the kurtosis gradient and outer slope of the velocity dispersion profile, but there is larger scatter in the $h_{4}$ gradient compared to observations. The TNG ETGs extend not as far as their MASSIVE Survey counterparts in the $\Delta h_{4}(R) / \Delta \mathrm{log} R$-$\gamma_{\mathrm{outer}}$ relation due to the less abundant sampling of the galaxy (Fig.~\ref{fig:3}) and halo (Fig.~\ref{fig:9}) masses in TNG100.}
    \label{fig:5}
\end{figure*}

\begin{figure}
	\includegraphics[width=\columnwidth]{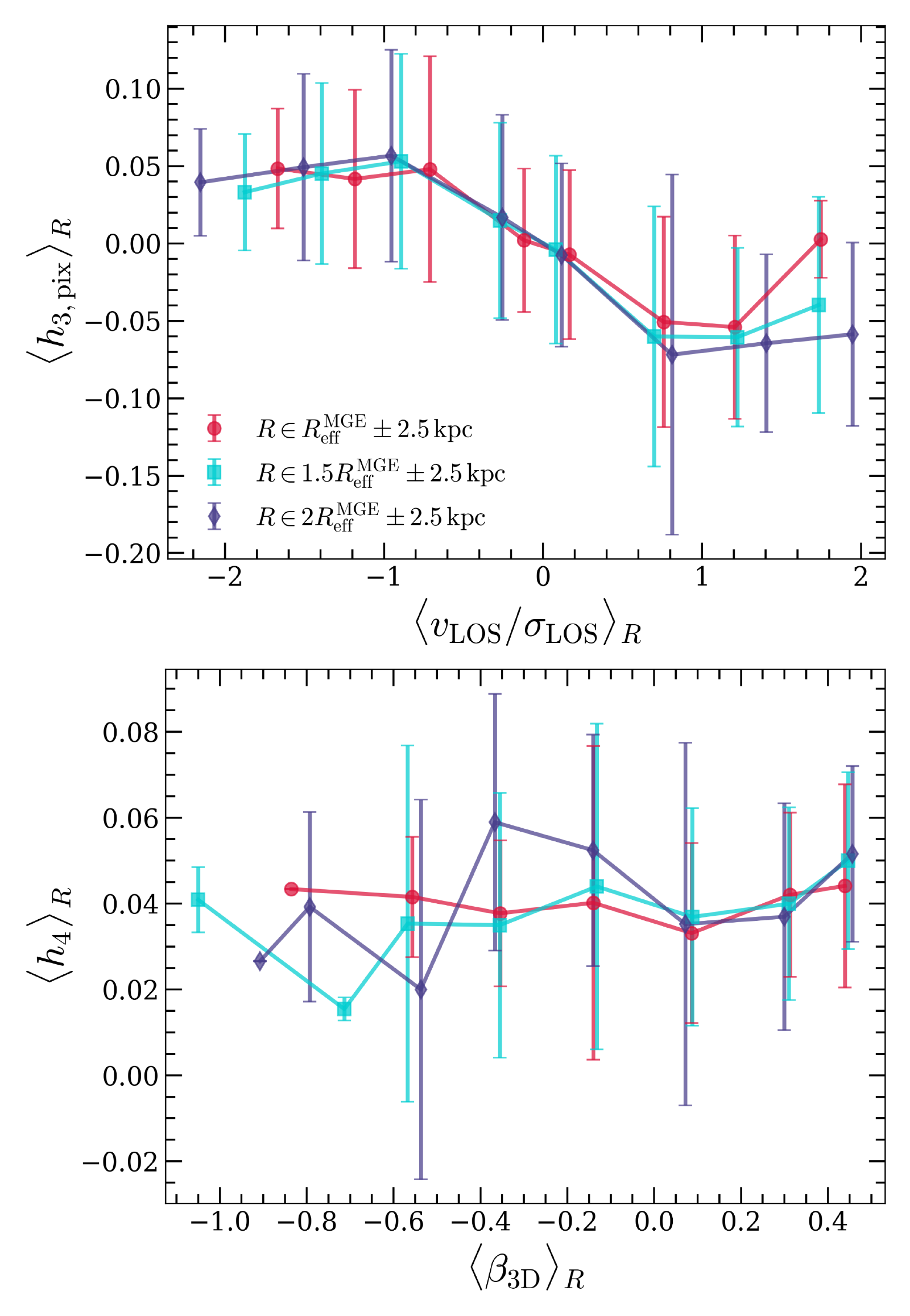}
    \caption{{\it Upper panel:} The correlation between the skewness $h_3$ and the rotation to dispersion ratio $v/\sigma$ at the pixel level measured around three different radii. {\it Lower panel:} The correlation between the average kurtosis $h_4$ and the spherical velocity anisotropy parameter $\beta_{\mathrm{3D}}$ measured in 3D apertures of different radii. Solid curves indicate the mean while the shaded regions/error bars represent the $1\sigma$ confidence intervals in both panels. } 
    \label{fig:6}
\end{figure}

In this section, we make direct comparison of the results of $h_{4}$ and $\gamma_{\mathrm{outer}}$ for IllustrisTNG ETGs with Figure 6 in \citet{2018MNRAS.473.5446V}, which is the pivotal comparison for this work. With the $h_{4}$ profiles for individual galaxies derived in the previous section, we compute the average kurtosis within the effective radii $\langle h_4 \rangle_{R_{\mathrm{eff}}^{\mathrm{MGE}}}$ with the SDSS $r$-band luminosity weighted value of $h_4$ in all spaxels/Voronoi bins projected within $R_{\mathrm{eff}}^{\mathrm{MGE}}$ of each galaxy. To derive the gradient of the kurtosis $\Delta h_{4}(R) / \Delta \mathrm{log} R$, we adopt the definitions in the observations~\citep{2017MNRAS.464..356V,2018MNRAS.473.5446V} and define:
\begin{equation}
    \label{eq:9}
    \frac{\Delta h_{4}(R)}{\Delta \mathrm{log} R} = \frac{h_{4}(1.5R_{\mathrm{eff}}^{\mathrm{MGE}}) - h_{4}(0.5R_{\mathrm{eff}}^{\mathrm{MGE}})}{\mathrm{log_{10}}(1.5R_{\mathrm{eff}}^{\mathrm{MGE}}) - \mathrm{log_{10}}(0.5R_{\mathrm{eff}}^{\mathrm{MGE}})},
\end{equation}
which is the luminosity-weighted average slope of $h_{4}$ within the logarithmic radial range of [$0.5R_{\mathrm{eff}}^{\mathrm{MGE}}$, $1.5R_{\mathrm{eff}}^{\mathrm{MGE}}$]. The spaxels/Voronoi bins included in the calculation of $h_4(R)$ at $1.5R_{\mathrm{eff}}^{\mathrm{MGE}}$ are taken from a thin annulus $\pm 0.5$ kpc that of $R$. Due to limited amount of spaxels available near the galactic center, we compute $h_4(R)$ at $0.5R_{\mathrm{eff}}^{\mathrm{MGE}}$ as the luminosity-weighted average $h_{4}$ for all spaxels within $R\leqslant 0.5 R_{\mathrm{eff}}^{\mathrm{MGE}}$. The results of the mean kurtosis $\langle h_4 \rangle_{R_{\mathrm{eff}}^{\mathrm{MGE}}}$ and the kurtosis gradient $\Delta h_{4}(R) / \Delta \mathrm{log} R$ versus the velocity dispersion profile outer slope $\gamma_{\mathrm{outer}}$ for our ETG sample are shown in Fig.~\ref{fig:5}. In both panels, observation values from \citet{2018MNRAS.473.5446V} are over-plotted for comparison to the simulation results. In the right panel, we also show the best-linear-fit comparison between the IllustrisTNG and MASSIVE $\Delta h_{4}(R) / \Delta \mathrm{log} R$-$\gamma_{\mathrm{outer}}$ relations. 

In the left panel of Fig.~\ref{fig:5}, the IllustrisTNG ETGs' average kurtosis are consistent with the range of $h_{4}$ values from observations. In the right panel of Fig.~\ref{fig:5}, although the kurtosis gradients of the simulated ETGs have larger scatter than the observed ETGs, it is remarkable that the simulation reproduces the key positive correlation between $\Delta h_{4}(R)/\Delta \mathrm{log} R$ and $\gamma_{\mathrm{outer}}$ as in Figure 6 of \citet{2018MNRAS.473.5446V}. The dotted-dashed blue line represents the best linear fit to the IllustrisTNG ETGs (slope $b = 0.092\pm0.035$, $p=0.009$, shaded region stands for $1\sigma$ confidence interval), and the dotted black line indicate the best linear fit to the MASSIVE ETGs (slope $b = 0.096\pm0.037$, $p=0.012$, see Fig. 6 in \citealt{2018MNRAS.473.5446V}). Thus, both the slope and correlation significance for IllustrisTNG ETGs are statistically consistent with the MASSIVE ETGs, which justifies the use of our mock ETG sample as a benchmark for understanding the origin of $\Delta h_{4}(R)/\Delta \mathrm{log} R$-$\gamma_{\mathrm{outer}}$ correlation in the outer kinematic structure of ETGs. 

To further check the consistency of higher order kinematic properties of our simulated ETGs with observations (e.g., Figure 10 in \citealt{2017MNRAS.464..356V}), we show the relation of the skewness $h_{3}$ versus the rotation to dispersion ratio $\langle v_{\mathrm{LOS}} \rangle / \sigma_{\mathrm{LOS}}$ in the upper panel of Fig.~\ref{fig:6}. Values of $h_{3}$ and $v/\sigma$ for individual spaxels/Voronoi bins from the different radial apertures (scaled to $R_{\mathrm{eff}}^{\mathrm{MGE}}$) in different galaxies are stacked together. Unlike $h_{4}$ which is the fourth velocity moment and has even parity, the reason why we keep track of the individual spaxel/Voronoi bin level $h_{3}$ information is due to its odd parity, otherwise projection in circular apertures would average out any underlying signal. The anti-correlation between $h_{3}$ and $v/\sigma$ seen for the simulated ETGs is consistent with the theoretical expectation that galaxies with substantial rotation (large $|v/\sigma|$) will have a large low velocity tail due to the projection effect of distant background and foreground stars along the line-of-sight, especially when the galaxy is viewed edge-on. The magnitude of $|h_{3}|\sim 0.1$ is prevalent from small ($\sim 1 R_{\mathrm{eff}}^{\mathrm{MGE}}$) to large ($\sim 2 R_{\mathrm{eff}}^{\mathrm{MGE}}$) radii, which also shows consistency with observations~\citep{2011MNRAS.414.2923K,2017MNRAS.464..356V}. Thus, both $h_{3}$ and $h_{4}$ measurements of the IllustrisTNG ETG sample from our mock IFU pipeline resembles that of the observed MASSIVE sample, indicating that physical interpretations (Section~\ref{sec:4}) of these kinematic structures analyzed from the simulation can be confidently applied to understanding the formation process of these features in real world ETGs. 

\subsection{De-correlated outer kurtosis and velocity anistropy}
\label{sec:3.4}

The core puzzle in \citet{2018MNRAS.473.5446V} that sparked the current study is the positive correlation between $\Delta h_{4}(R)/\Delta \mathrm{log} R$ and $\gamma_{\mathrm{outer}}$. If an intrinsically flat velocity dispersion profile (assuming no ordered rotation in an ideal ETG, $v/\sigma$=0) is affected by the projection effects of velocity anisotropy at the outskirts, we expect to find radial anisotropy driving positive $h_{4}$ and `dropping' $\gamma_{\mathrm{outer}}$ (vice versa for the tangential case), producing an anti-correlation of $\Delta h_{4}(R)/\Delta \mathrm{log} R$ and $\gamma_{\mathrm{outer}}$~\citep{1998MNRAS.295..197G}. This theoretical expectation is exactly opposite to what was observed in \citet{2018MNRAS.473.5446V}. A crucial assumption of this prediction is that they were evaluated under fixed density profiles~\citep{1993MNRAS.265..213G}, typically comprising a singular isothermal sphere halo potential (fixed $\gamma^{\prime}=2$) and a spherical stellar distribution that have a steeper density profile (fixed $\gamma^{\prime}=3$ or $4$), which naturally leads to \citet{2018MNRAS.473.5446V} proposing variations in the total density profile as an explanation.

Before presenting the connection between the outer kinematic structure and the density profile, we show in the lower panel of Fig.~\ref{fig:6} how the velocity anisotropy ($\beta_{\mathrm{3D}}$) correlate with $h_{4}$ in our ETGs. The spherical velocity anisotropy is defined as~\citep{2008gady.book.....B}:
\begin{equation}
\label{eq:10}
\beta_{\mathrm{3D}} = 1 - \frac{\sigma_{\mathrm{\phi}}^{2} + \sigma_{\mathrm{\theta}}^{2}}{2\sigma_{\mathrm{r}}^{2}}\,,
\end{equation}
where $\sigma_{\mathrm{\phi}}$, $\sigma_{\mathrm{\theta}}$, and $\sigma_{\mathrm{r}}$ are the velocity dispersion in the azimuthal, polar, and radial directions. As shown in Fig.~\ref{fig:6}, there is almost no correlation between $\beta_{\mathrm{3D}}$ and $h_{4}$ averaged in across different aperture sizes. Particularly, we find a positive $h_4$ (median $\sim 0.04$) even for tangential anisotropy ($\beta<0$), and only a slight positive trend consistent across different radii at $\langle \beta_{\mathrm{3D}} \rangle_{R} \gtrsim 0.1$. This suggests that positive $h_{4}$ towards the galactic outskirts is unlikely driven by systematic radial velocity anisotropy prevalent in ETGs.

In concordance with the conjecture in~\citet{2018MNRAS.473.5446V}, we elaborate in Section~\ref{sec:4.1} that the positive $\Delta h_{4}(R)/\Delta \mathrm{log} R$-$\gamma_{\mathrm{outer}}$ correlation is in fact driven by the variation of their total density profiles across our ETG sample. The variations in the outer kurtosis and velocity anisotropy also show hints to be driven by different routes of stellar assembly (Section~\ref{sec:4.2}). The de-correlated behavior between the outer $h_{4}$ and $\beta_{\mathrm{3D}}$ ($R \geqslant R_{\mathrm{eff}}^{\mathrm{MGE}}$) shown in Fig.~\ref{fig:6} can then be accounted for by the dominant effect from gradients in circular velocity~\citep{1993MNRAS.265..213G,2005A&A...432..411B}, where the effect of varying density profiles can overwhelm and randomize the rather weak correlation between $h_{4}$ and velocity anisotropy.

\section{Linking kinematic structure with density profile, assembly history and environment}
\label{sec:4}

As shown in the previous section, the IllustrisTNG ETGs have consistent outer slopes of the velocity dispersion profile ($\gamma_{\mathrm{outer}}$) and higher order velocity moments ($h_{3}$, $h_{4}$) compared to observations. The simulated ETGs also reproduce the observed positive correlation between $\Delta h_{4}(R)/\Delta \mathrm{log} R$ and $\gamma_{\mathrm{outer}}$, and that velocity anisotropy is insufficient to explain the formation of this trend. In this section, we explore the validity of circular velocity gradients as a driving factor of this trend of following the hypothesis in~\citet{2018MNRAS.473.5446V}. We also explore the connection of their kinematic structure to their stellar assembly and environment. 

\begin{figure*}
	\includegraphics[width=2\columnwidth]{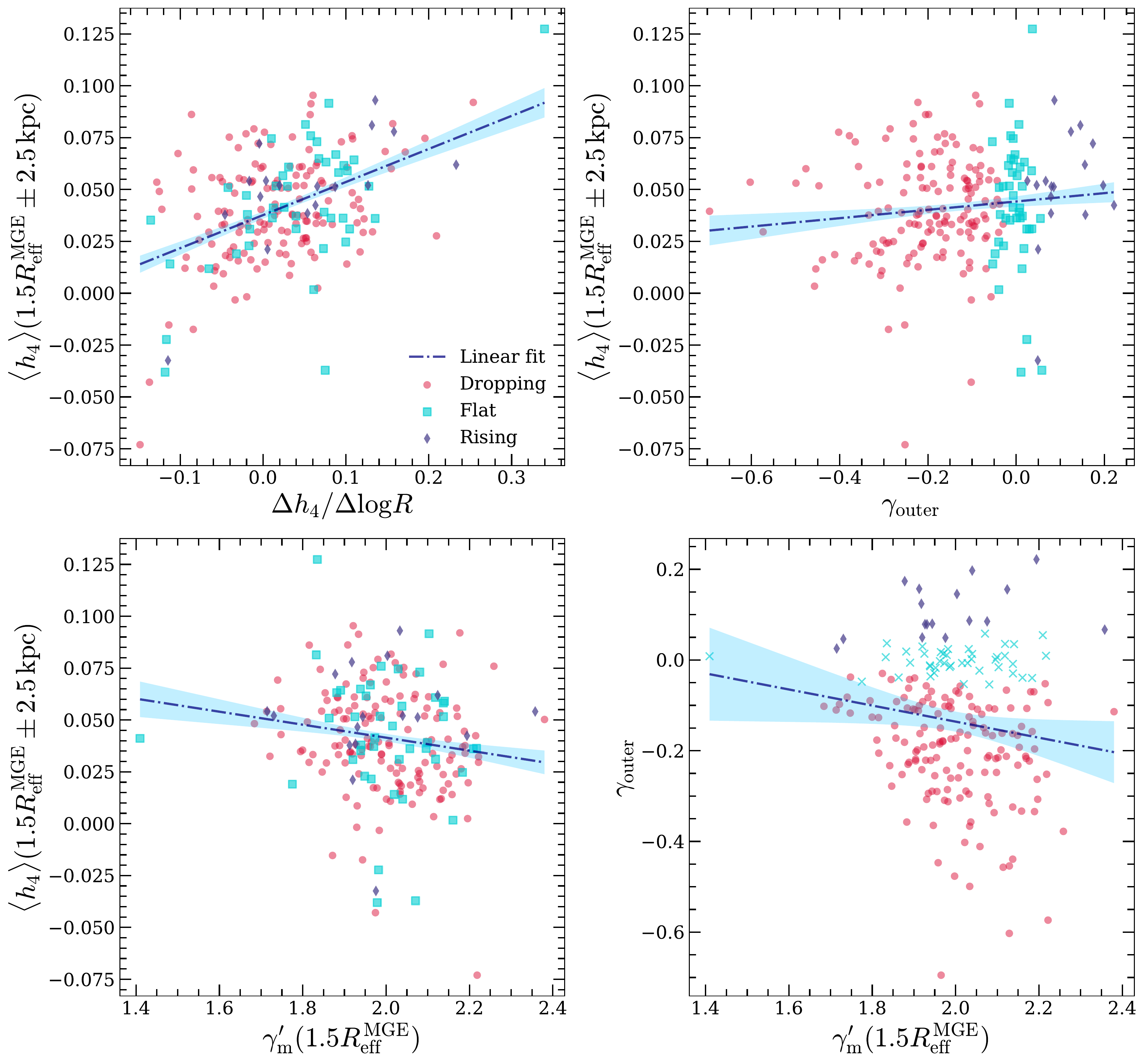}
    \caption{{\it Upper left:} the relation between the $h_{4}$ gradient and the outer $h_{4}$ values measure at around $1.5R_{\mathrm{eff}}^{\mathrm{MGE}}$. {\it Upper right:} the relation between the outer $h_{4}$ values and the outer velocity dispersion profile slopes $\gamma_{\mathrm{outer}}$. {\it Lower left:} the relation between outer $h_{4}$ and the mass-weighted density profile slope $\gamma^{\prime}_{\mathrm{m}}$ (Equation~\ref{eq:11}) measured at $1.5R_{\mathrm{eff}}^{\mathrm{MGE}}$ . {\it Lower right:} the relation between the velocity dispersion profile outer slope and the density profile slope. In all panels, The different ETGs with rising, flat, and dropping $\gamma_{\mathrm{outer}}$ are indicated by the different-colored scattered dots, following Fig.~\ref{fig:4}. The blue dotted-dashed lines and shaded regions stand for the best linear fit and their [$16\%$, $84\%$] confidence intervals for each of the correlations.}
    \label{fig:7}
\end{figure*}

\subsection{Correlation with density profile}
\label{sec:4.1}

Before proceeding into the details of exploring the density profiles, we make a change in the quantity characterizing the non-Gaussian outer kinematic structure from the kurtosis gradient $\Delta h_{4}(R)/\Delta \mathrm{log} R$ to the outer kurtosis $\langle h_{4} \rangle$ (measured at $1.5 R_{\mathrm{eff}}^{\mathrm{MGE}} \pm 2.5\,\mathrm{kpc}$), which is the luminosity-weighted $h_{4}$ for spaxels/Voronoi bins within a 5 kpc annulus around $1.5 R_{\mathrm{eff}}^{\mathrm{MGE}}$. The motivation for our choice is two fold: on the one hand, since we focus on the correlation with $\gamma_{\mathrm{outer}}$, we would prefer to isolate the contribution from outer $h_{4}$ alone; on the other hand, this change of variable can also avoid the systematics in the stellar velocity distribution at small radii pertinent to resolution effects (core softening) and feedback physics (dark matter fraction) as discussed in Section~\ref{sec:3.2}. In the upper left panel of Fig.~\ref{fig:7}, we show $\Delta h_{4}(R)/\Delta \mathrm{log} R$ versus the outer $h_{4}$ for the IllustrisTNG ETGs. The best linear fit of the correlation has slope $b=0.159\pm0.022$, $p=5.12\times 10^{-12}$, and Pearson correlation coefficient $r = 0.456$, justifying that the outer $h_{4}$ is a decent representative of the kurtosis gradient. Following that, we show in the upper right panel of Fig.~\ref{fig:7} that a positive correlation (slope $b=0.020\pm0.012$, $p=0.104$) also exists between the outer $h_{4}$ and the outer velocity dispersion profile slope $\gamma_{\mathrm{outer}}$. Combining these two panels, we anchor our following discussion regarding the link to density profiles and assembly history on the outer $h_{4}$. 

The power-law slope of the total density profile, $\gamma_{\rm PL}^{\prime}$ (same as $\gamma^{\prime}$ in Equation~\ref{eq:1}, renaming to avoid confusion with the velocity dispersion profiles' inner and outer slopes) reflects the underlying gradient in circular velocity (interchangeable with velocity dispersion in the isotropic case, Equation~\ref{eq:2}) such that $\gamma^{\prime}_{\mathrm{PL}} > 2$ corresponds to a `dropping' circular velocity profile, $\gamma^{\prime}_{\mathrm{PL}} < 2$ a `rising' circular velocity profile, and a flat circular velocity profile in the exact isothermal case ($\gamma^{\prime}_{\mathrm{PL}} = 2$). In Paper I and II, this parameter was obtained by a linear fit to the combined $\log\rho$-$\log r$ profile of dark matter, stars, and gas in 100 logarithmic radial bins from $0.4R_{\mathrm{eff}}$ to $4R_{\mathrm{eff}}$, where $R_{\mathrm{eff}}$ is the projected stellar half mass radius. To study the total density profile at the outskirts of these ETGs, we approximate the power-law slope with the `mass-weighted' logarithmic density slope~\citep{2014MNRAS.438.3594D,2020MNRAS.491.5188W}:
\begin{equation}
    \label{eq:11}
    \gamma^{\prime}_{\mathrm{m}}(R) =
    \frac{-1}{M(<R)}\int_{0}^{R} \gamma^{\prime}_{\mathrm{PL}}(r) 4\pi r^{2} \rho(r) dr = 3 - \frac{4\pi R^{3}\rho(R)}{M(<R)},
\end{equation}
where $M(<R)$ is the mass enclosed within a 3D spherical aperture with radius $R$, and $\rho(R)$ is the average local density at radius $R$. We evaluate the mass-weighted slope $\gamma^{\prime}_{\mathrm{m}}$ at $1.5 R_{\mathrm{eff}}^{\mathrm{MGE}}$ (local density $\rho(R)$ calculated in a thin shell $\pm2\%$ at this radius) to compare with the outer $h_{4}$ and $\gamma_{\mathrm{outer}}$. The correlations along with the best linear fit results are shown in the bottom panels in Fig.~\ref{fig:7}.

\begin{figure*}
	\includegraphics[width=2\columnwidth]{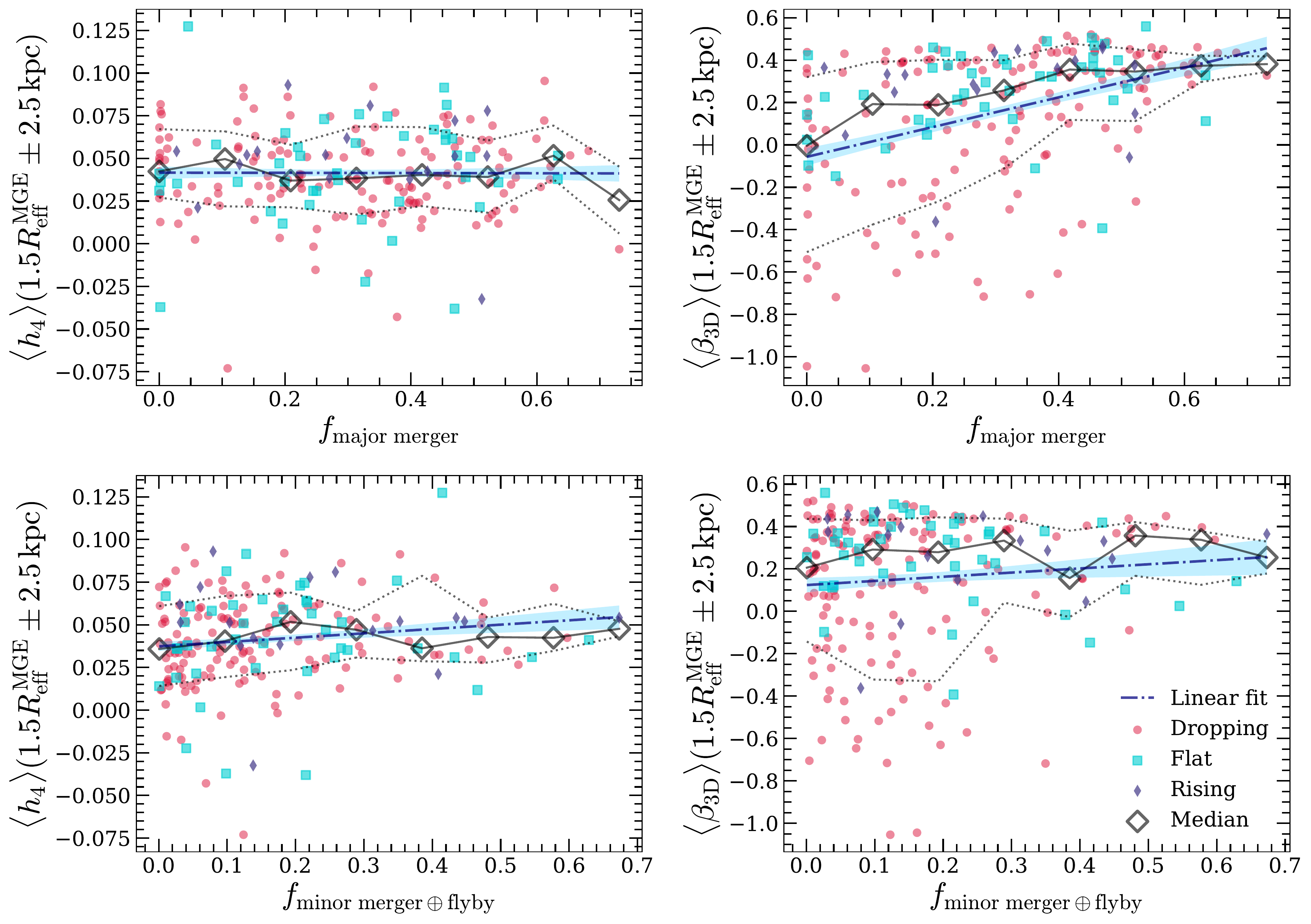}
    \caption{{\it Left column:} The relation between the outer $h_{4}$ values and stellar assembly contributions. Major (minor) mergers are defined as mergers having stellar mass ratio larger (smaller) than 0.25. {\it Right column:} The relation between the outer $\beta_{\mathrm{3D}}$ and stellar assembly contributions. In each column, the upper panel shows the link to the fraction of stellar mass accreted in major mergers, while the lower panel shows the link to the fraction of stellar mass accreted in minor perturbations, i.e. minor mergers and flyby interactions. {\it The key takeaway} is that the outer $\langle h_4 \rangle$ correlates with minor perturbations but not major mergers, while the velocity anisotropy $\langle \beta_{\mathrm{3D}} \rangle$ correlates more strongly with major mergers. The three types of ETGs with rising, flat, and dropping outer velocity dispersion profiles are shown by the scattered dots. The open markers with solid lines (dashed lines) denote the median ($68\%$ scatter) of the rising, dropping, and flat $\gamma_{\mathrm{outer}}$ ETGs combined. The blue dotted--dashed lines and shaded regions stand for the best linear fit and their [$16\%$, $84\%$] confidence intervals (please see text in Section~\ref{sec:4.2} for best-fit slopes and $p$-values.)}
    \label{fig:8}
\end{figure*}

Intriguingly, both the outer $h_{4}$ (bottom left panel) and the velocity dispersion profile outer slope $\gamma_{\mathrm{outer}}$ (bottom right panel) are anti-correlated with the outer mass-weighted density profile slope at $1.5R_{\mathrm{eff}}^{\mathrm{MGE}}$. Apparently, there is noticeable scatter in both relations which lead to the scatter seen in the $\Delta h_{4}(R)/\Delta \mathrm{log} R$-$\gamma_{\mathrm{outer}}$ correlation in Fig.~\ref{fig:5}. Nevertheless, both of these negative correlations are statistically significant, with $p=0.028$ for the outer $h_{4}$-$\gamma^{\prime}_{\mathrm{m}}$ correlation (slope $b=-0.031\pm0.014$) and $p=0.027$ for the $\gamma_{\mathrm{outer}}$-$\gamma^{\prime}_{\mathrm{m}}$ correlation (slope $b=-0.177\pm0.079$). With the velocity anisotropy ruled out as a plausible explanation as discussed in Section~\ref{sec:3.4}, these two correlations suggest that it is indeed the systematic variation of the total density profile being shallower than isothermal ($\gamma^{\prime}_{\mathrm{m}}<2$) that leads to `rising' gradients in the velocity dispersion profiles ($\gamma_{\mathrm{outer}}>0$) and produces positive outer $h_{4}$ in these ETGs (vice versa for the steeper than isothermal case where $\gamma^{\prime}_{\mathrm{m}}>2$), giving rise to an overall positive correlation between $\Delta h_{4}(R)/\Delta \mathrm{log} R$ and $\gamma_{\mathrm{outer}}$. The physical interpretation behind this correlation is that ongoing or past remnants of minor mergers producing positive outer $h_4$ while simultaneously driving the total density profile shallower than isothermal (Section~\ref{sec:4.2}). Moreover, combined with our finding in Paper I that there is no significant correlation between the density profile slope and velocity anisotropy (see Figure 8 in \citealt{2020MNRAS.491.5188W}), having the main driving factor of a non-zero $h_{4}$ being a deviation from an isothermal density profile allows for randomness between $h_{4}$ and $\beta_{\mathrm{3D}}$, leading to the de-correlated $h_{4}$-$\beta_{\mathrm{3D}}$ trend as seen in the bottom panel of Fig.~\ref{fig:6}. 

\subsection{Correlation with stellar assembly history}
\label{sec:4.2}

Galaxy mergers play an important role in the non-dissipative (dry) evolution of ETGs at low redshift~\citep{2009ApJ...703.1531N,2009ApJ...706L..86N,2013ApJ...766...71R,2016MNRAS.456.1030W}. In Paper I we demonstrated that the stellar assembly history of ETGs parameterized by the fraction of in-situ-formed stars ($f_{\mathrm{in-situ}}$) positively correlates with the total density profile power-law slope within $4R_{\mathrm{eff}}$ (see Figure 10 in \citealt{2020MNRAS.491.5188W}). In Paper II we have shown that gas-poor mergers at low redshift ($z\lesssim 0.5$), both major and minor mergers alike, work in coordination to reduce  $\gamma^{\prime}$ of the total density profile and establish the positive $f_{\mathrm{in-situ}}-\gamma^{\prime}$ correlation (see Figure 7 in \citealt{2019MNRAS.490.5722W}). Motivated by these finding and the positive relation of $h_{4}$ at $1.5R_{\mathrm{eff}}^{\mathrm{MGE}}$ and $\gamma^{\prime}_{\mathrm{m}}$ as shown above in Fig.~\ref{fig:7}, we expect to also find a correlation between the outer $h_{4}$ and the stellar assembly history of these ETGs. 

In Fig.~\ref{fig:8}, we present the relations between the outer $h_{4}$ (left column) and outer $\beta_{\mathrm{3D}}$ (right column) measured at $1.5R_{\mathrm{eff}}^{\mathrm{MGE}}$ as a function of the mass fraction of stars in these ETGs accreted either from major mergers (upper row) or minor perturbations combining the contribution from minor mergers and flyby interactions (lower row). Major (minor) mergers are defined as mergers into the main progenitors of our $z=0$ ETGs having stellar mass ratios larger (smaller) than $1/4$. The solid line and open markers denote the median of the combined sample including the rising, dropping, and flat outer velocity dispersion profile ETGs, while the dashed lines represent the [$16\%$, $84\%$] range ($1\,\sigma$) of the distributions. The outer $h_{4}$ shows almost no correlation with the stellar constituent from major mergers (best linear fit slope $b = -0.0006\pm0.0103$, $p = 0.9545$, statistically insignificant). Intriguingly, the outer $h_{4}$ demonstrates a positive correlation (best linear fit slope $b = 0.0254\pm0.0128$, $p = 0.0474$, statistically significant) with the stellar assembly contribution from minor perturbations at $f_{\mathrm{minor\ merger\oplus flyby}}\lesssim 0.3$. This correlation is shaped by the majority of galaxies in our sample which demonstrates consistent increase in the median as well as the $1\,\sigma$ region. We also notice that there is a slight decrease in outer $h_{4}$ at $f_{\mathrm{minor\ merger\oplus flyby}}\sim 0.4$, beyond which $h_4$ remains largely constant. This fluctuation in the outer $h_4$ at high minor perturbation fractions could be due to the limited sample size at the massive end.

Nevertheless, the outer 3D velocity anisotropy shows positive correlations with stellar assembly through both major mergers and minor perturbations, with both relations having large scatter as well. The main difference regarding these two correlations is that major mergers seem to drive a more significant ($\sim 0.4$ versus $\sim 0.1$) and steady (monotonic versus fluctuating) increase in the outer $\beta_{\mathrm{3D}}$ compared to minor perturbations. This is also evident from the steeper slope of the linear fit between $\langle \beta_{\mathrm{3D}} \rangle$ and $f_{\mathrm{major\ merger}}$ (slope $b = 0.702\pm0.115$, $p = 5.83\times 10^{-9}$, very significant) versus $f_{\mathrm{minor\ merger\oplus flyby}}$ (slope $b = 0.196\pm0.157$, $p = 0.213$, poor statistical significance). Hence, the positive outer $h_{4}$ in ETGs tends to be built up by minor mergers and galaxy interactions (at least for the majority of galaxies with $f_{\mathrm{minor\ merger\oplus flyby}}\lesssim 0.4$), but the radial anisotropy of stars ($\beta_{\mathrm{3D}}$) at the galaxy outskirts is mainly driven by major mergers, resulting in an almost decoupled $h_{4}$-$\beta_{\mathrm{3D}}$ relation as seen in Fig.~\ref{fig:6}.

Although the four sets of relations shown above all possess scatter, their differences hint that the seemingly decoupled outer $h_{4}$ and $\beta_{\mathrm{3D}}$ (Fig.~\ref{fig:6}) trends originate from the different impacts of major and minor perturbations during the formation process of ETGs. Major mergers can create cuspy (wet merger) or cored (dry merger) central stellar densities depending on the gas fraction of the merger~\citep{2008ApJ...679..156H,2009ApJS..181..135H,2009ApJS..182..216K,2009ApJS..181..486H,2009ApJ...691.1168H,2019MNRAS.487.5416T}, modifying the central density profile. Major mergers can also drastically change the morphology and dynamics of a massive galaxy from being tangentially biased disk galaxies (late-type) to radial-orbit-dominated ETGs~\citep{2009ApJS..182..216K,2012ApJS..198....2K,2017MNRAS.467.3083R}, which explains the positive correlation between $\beta_{\mathrm{3D}}$ and $f_{\mathrm{major-merger}}$. Although in isolated galaxy merger simulations, a bump in the velocity dispersion profile can also be created at radii $>R_{\mathrm{eff}}$ via major mergers ~\citep{2014ApJ...783L..32S}, simultaneous minor mergers could counter-act that effect in a cosmological setting~($1\leqslant z \leqslant 2$, \citealt{2019MNRAS.490.5722W}). In addition, minor mergers contribute significantly to the growth of ETGs compared to major mergers~\citep{2008MNRAS.384....2G}, perpetuate the inside-out growth of extensive stellar halos and modify stellar velocity distributions of ETGs at large radii~\citep{2007MNRAS.377....2M,2009MNRAS.395.1491B,2011MNRAS.417..845K,2012MNRAS.425.3119H,2013MNRAS.429.2924H,2021A&A...647A..95P,2021MNRAS.504.4923D}. 

By visually inspecting the kinematic maps of $h_{3}$ and $h_{4}$ for individual galaxies, we find that at the locations where $h_{3}$ and $h_{4}$ locally peak, it is also a common location to find a coinciding overdensity in the projected luminosity map (see Appendix~\ref{sec:App_B} for examples). These overdensities reflect the transient infall phases of accreted satellite galaxies on to the host ETGs, before they are tidally-disrupted and identified as minor mergers by \textsc{Sublink}. The corresponding LOS velocity profile in the spaxels/Voronoi-bins that cut through these satellites show peaks that are phase from the mean velocity of the galaxy, which leads to large $h_{3}$ and $h_{4}$ values in those spaxels/Voronoi-bins. Eventually, the cumulative effect of these minor perturbations (which are predominantly dry mergers at low redshift, see Figure 7~\citealt{2019MNRAS.490.5722W} for simulation evidence and \citealt{2021MNRAS.506.3691D} for latest dynamical modeling constraints from VLT-MUSE) not only leads to positive outer $h_{4}$, but also drives the outer density profile to be shallower than isothermal ($\gamma^{\prime}_{\mathrm{m}}$), establishing the positive correlation between $\Delta h_{4}(R)/\Delta \mathrm{log} R$ and $\gamma_{\mathrm{outer}}$. Combined with the aforementioned impact of minor mergers on the formation of ETGs, the formation of the non-Gaussian outer kinematic structure in early-type galaxies is a natural consequence of hierarchical structure formation, and it co-evolves with the outer density profile mainly via contributions from minor perturbations. As we will discuss in the following, these trends are also consistent with the results from the probes of galaxy (halo) environment.

\subsection{Correlation with environment}
\label{sec:4.3}

\begin{figure*}[t!]
	\includegraphics[width=2\columnwidth]{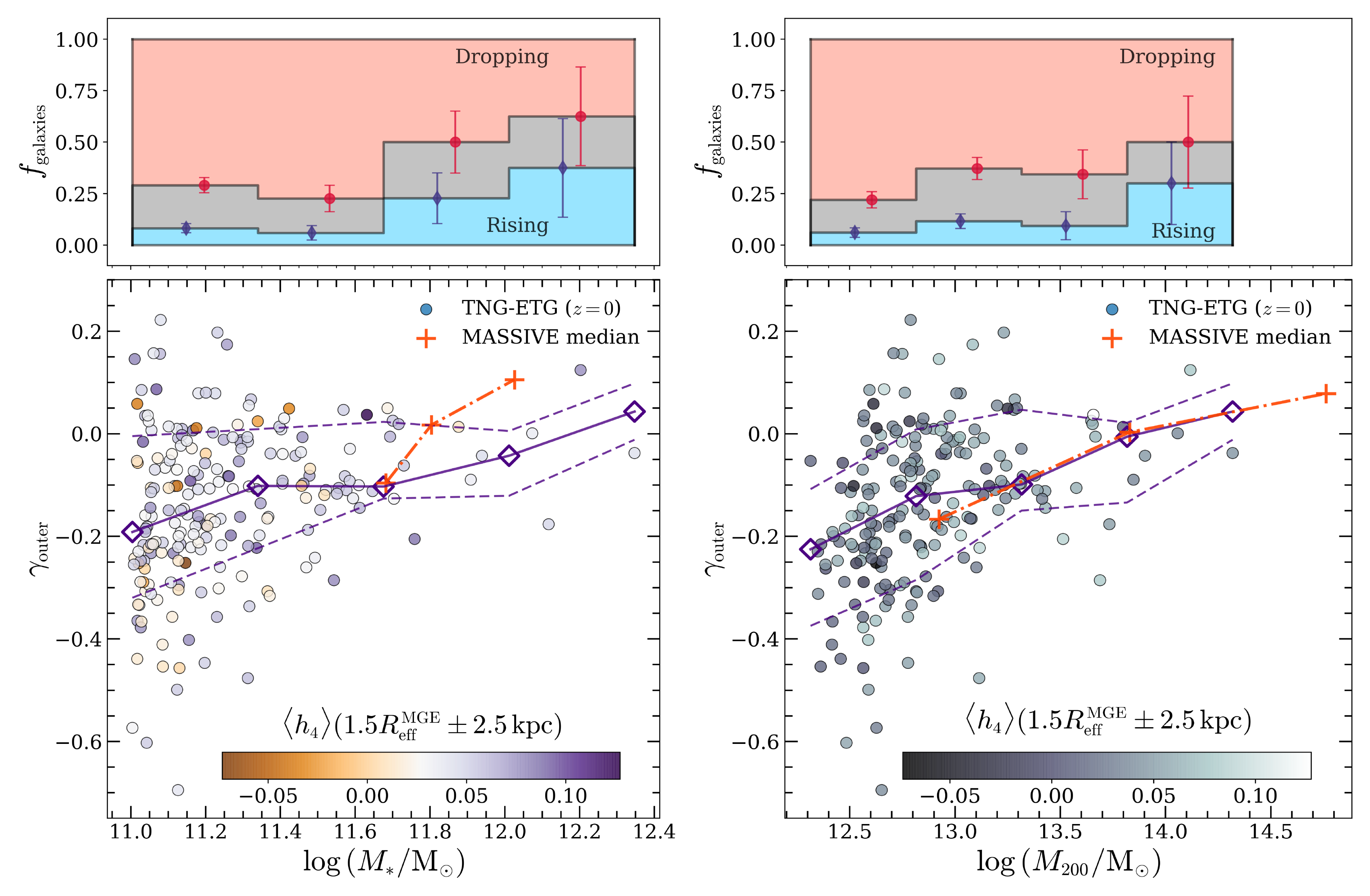}
    \caption{{\it Left panel:} The correlation between the outer velocity dispersion profile slope ($\gamma_{\mathrm{outer}}$) and the total stellar mass ($M_{\ast}$) of our simulated ETGs, color-coded by their outer $h_{4}$ values measured at around $1.5 R_{\mathrm{eff}}^{\mathrm{MGE}}$. {\it Right panel:} The correlation between $\gamma_{\mathrm{outer}}$ and the host halo mass ($M_{200}$) of our simulated ETGs, color-coded by their outer $h_{4}$ values measured at around $1.5 R_{\mathrm{eff}}^{\mathrm{MGE}}$. The open markers with solid lines in both panels denote the median, while the dashed lines represent the [$16\%$, $84\%$] of the distribution for $\gamma_{\mathrm{outer}}$ in different stellar mass and halo mass bins. The red crossed with dotted-dashed lines are median values for the same correlations from \citet{2018MNRAS.473.5446V}. The upper side plots show the fraction of galaxies with rising (blue), flat (grey), and dropping (red) outer dispersion profiles in each stellar/halo mass bin. The error bars in each bin stand for the $68\%$ Wilson score confidence intervals.}
    \label{fig:9}
\end{figure*}

Another interesting feature related to the $\Delta h_{4}(R)/\Delta \mathrm{log} R$-$\gamma_{\mathrm{outer}}$ found in ~\citet{2018MNRAS.473.5446V} is that there is an increasing fraction of ETGs with `rising' outer velocity dispersion profiles living in denser environments and more massive halos. Again, the authors speculate that this is also induced by the systematic variation of density profiles instead of velocity anisotropy across different ETGs. Since the environmental measures $\nu_{10}$ and $\delta_{g}$ in \citet{2017MNRAS.464..356V,2018MNRAS.473.5446V} are based on the galaxy K-band luminosity weighted on the MASSIVE Survey limiting magnitude and completeness, we do not attempt to perform mock observations of our simulated ETGs based on raw SDSS $K$-band magnitude output by IllustrisTNG. Hence, we focus on comparing the total stellar ($M_{\ast}$) and host dark matter halo masses ($M_{200}$, mass contained in the radius where the mean density is $200\times$ the universal critical density) of our ETG sample to the trends observed in the MASSIVE Survey. The stellar and halo masses serve as sensible proxies of the ETG environment as all targets in MASSIVE are central galaxies and we follow the same selection for our mock IllustrisTNG ETGs.

In Fig.~\ref{fig:9}, we show $\gamma_{\mathrm{outer}}$, as well as the fraction of ETGs having `rising', `flat', or `dropping' velocity dispersion profile outer slopes, against their stellar (left panel) and halo masses (right panel). The scattered dots are individual galaxies colored by their outer $h_{4}$ values, the solid curve with diamond markers denotes the binned median, and the dashed lines represent the [$16\%$, $84\%$] interval of the distribution within each mass bin. We see that $\gamma_{\mathrm{outer}}$ increases with both increasing stellar and halo masses, in agreement with the trend of MASSIVE median values (red crosses). We notice that the MASSIVE ETGs (89 galaxies, \citealt{2018MNRAS.473.5446V}, also see their Figures 7, 8, and 9 which also shows large scatter in $\gamma_{\mathrm{outer}}$) have a slightly steeper relation with $M_{\ast}$, possibly due to a more complete sample of heavy galaxies with $\log_{10}(M_{\ast}/\mathrm{M_{\astrosun}})\geqslant 11.6$ compared to the IllustrisTNG ETG sample (18 galaxies). The same trend is reflected in the upper side plots for the fraction of galaxies with different velocity dispersion profile outer slopes. The general trend in both columns indicate that higher fractions of galaxies will have `rising' $\gamma_{\mathrm{outer}}$ with increasing galaxy and halo mass (the error bars in the upper panels are the Wilson score $68\%$ confidence intervals, \citealt{Wilson1927}). 

Recasting on the findings in Paper I~\citep{2020MNRAS.491.5188W}, the power-law slope of the total density profile anti-correlating with the total stellar mass (Figure 4) is driven by the variation in the inner dark matter density profile (Figure 17), and hence anti-correlating with the halo mass (Figure 15). Given that the variation in $\gamma_{\mathrm{outer}}$ is mainly driven by a negative correlation with the density profile (lower right panel in Fig.~\ref{fig:7}), the findings in Paper I self-consistently predict the trends seen in Fig.~\ref{fig:9}, once again confirming the conjecture for the observed phenomena in~\citet{2018MNRAS.473.5446V}. The color index in the figure shows that the outer $h_{4}$ also tend to be higher in ETGs with higher stellar and halo mass, consistent with the positive $\Delta h_{4}(R)/\Delta \mathrm{log} R$-$\gamma_{\mathrm{outer}}$ relation. 

The physical interpretation for the emergence of the $\gamma_{\mathrm{outer}}$-mass correlation also fits in to the multi-phase formation path of near-isothermal ETG density profiles in Paper II~\citep{2019MNRAS.490.5722W}: heavier ETGs are quenched earlier by AGN feedback (Figures 4 and 8) and enter earlier into the non-dissipative evolution phase dominated by gas-poor mergers at lower redshift $z\lesssim 2$ (Figure 7), where the total density profile progressively evolves shallower, and more massive galaxies living in more massive halos (denser environments) experience more mergers. Although ETGs go through both major and minor mergers, with previous studies suggesting that major mergers start to dominate the ex-situ stellar population at stellar masses $\log_{10} (M_{\ast}/\mathrm{M_{\astrosun}}) \gtrsim 11.2$ (e.g., Fig. 10 in \citealt{2019MNRAS.487.5416T}), which is also the case for our ETG sample with median $f_{\mathrm{major\ merger}}$ = 0.31 and median $f_{\mathrm{minor\ merger\oplus flyby}}=0.12$, the stellar mass fraction from both minor and major mergers increase with $M_{\ast}$. Since the outer $h_4$ is mainly driven by minor perturbations instead of major mergers (Section~\ref{sec:4.2}, lower left panel in Fig.~\ref{fig:8}), the mass trend of $f_{\mathrm{minor\ merger\oplus flyby}}$ alone dictates the mass trend of $\gamma_{\mathrm{outer}}$ and outer $h_4$. This eventually creates the positive correlation between $\gamma_{\mathrm{outer}}$ and the kurtosis gradient through the mass-dependent variation of the total density profile, indicating that the formation of the outer (non-Gaussian) kinematic structure in ETGs stands as a natural consequence of hierarchical galaxy-halo assembly. 

\section{Conclusions and Outlook}
\label{sec:5}


We study the relation between the kinematic structure and density profiles of 207 massive ($\log_{10}\left( M_{\ast}/\mathrm{M_{\astrosun}} \right) \in [11.0, 12.4]$, non-major-merger, Fig.~\ref{fig:2}) early-type-galaxies selected from the IllustrisTNG Simulations~\citep{2018MNRAS.480.5113M,2018MNRAS.477.1206N,2018MNRAS.475..624N,2018MNRAS.475..648P,2018MNRAS.475..676S}. We produce mock integral-field-unit spectroscopic observations (Fig.~\ref{fig:1}) for these ETGs in random projections and compute the second (velocity dispersion $\sigma$), third (skewness $h_{3}$), and fourth (kurtosis $h_{4}$) velocity moments of the projected stellar velocity profiles (Fig.\ref{fig:3}). The outer slope ($\gamma_{\mathrm{outer}}$ ) of the LOSVD profiles (Fig.~\ref{fig:4}) and the key positive correlation between $\gamma_{\mathrm{outer}}$ and the kurtosis gradient ($\Delta h_{4}(R)/\Delta \mathrm{log} R$, Fig.~\ref{fig:5}) for the simulated ETGs are consistent with that found in the MASSIVE Survey~\citep{2014ApJ...795..158M,2017MNRAS.464..356V,2018MNRAS.473.5446V}. This realistic mock ETG sample serves as a benchmark for disentangling the correlation between the outer non-Gaussian kinematic features and velocity dispersion profiles in observed ETGs from the MASSIVE Survey~\citep{2018MNRAS.473.5446V}. In particular, we find:
\begin{itemize}
    
    \item The average velocity anisotropy and $h_{4}$ show almost \emph{no} correlation with the 3D velocity anisotropy $\beta_{\mathrm{3D}}$ at radii $\geqslant R_{\mathrm{eff}}^{\mathrm{MGE}}$ (Fig.~\ref{fig:6} lower panel), indicating that positive outer $h_{4}$ values is unlikely to be  produced by stellar velocity anisotropy under fixed density profiles~\citep{1993MNRAS.265..213G}.
    
    \item The outer $h_{4}$ measured at $\sim 1.5 R_{\mathrm{eff}}^{\mathrm{MGE}}$ is a good proxy for the kurtosis gradient and also positively correlates with $\gamma_{\mathrm{outer}}$ (Fig.~\ref{fig:7} upper row). The mass-weighted slope of the total density profile at that radius negatively correlates with both the outer $h_{4}$ and $\gamma_{\mathrm{outer}}$ (Fig.~\ref{fig:7} lower row). This justifies the conjecture~\citep{2018MNRAS.473.5446V} that systematic variations around an exact isothermal profile drive the positive correlation between $\Delta h_{4}(R)/\Delta \mathrm{log} R$ and $\gamma_{\mathrm{outer}}$. 
    
    \item The outer $h_{4}$ positively correlates with the mass fraction of stars accreted via {\it minor} perturbations (minor mergers and flyby interactions), while the outer velocity anisotropy correlates more significantly with the stellar mass fraction from {\it major} mergers (Fig.~\ref{fig:8}), resulting in a seemingly decoupled $h_{4}$--$\beta_{\mathrm{3D}}$ relation at the ETG outskirts (Fig.~\ref{fig:6}).
    
    \item The values of $\gamma_{\mathrm{outer}}$, outer $h_{4}$, and the fraction of ETGs with `rising' outer velocity dispersion profiles increase with increasing stellar and halo masses. This is mainly driven by the increase in minor merger fractions with increasing ETG stellar mass affecting the outer kinematic structure and density profile simultaneously~(Fig.~\ref{fig:9}).
    
\end{itemize}

As an augmentation to the detailed studies of the ETG density profiles in Paper I and II, this work generalizes those findings and links them with the outer kinematic structure in these galaxies. Our finding that the outer $h_{4}$ correlates with minor mergers and the density profile, while the outer $\beta_{\mathrm{3D}}$ correlates with major mergers, also explains the apparent stochasticity between the total density profile and stellar velocity anisotropy of ETGs found in Figure 8 of Paper I. The outer $h_{4}$--minor merger correlation is also consistent with our previous finding: in systems where the stellar assembly in ETGs is dominated by minor perturbations at low redshift ($z<0.5$),  a higher accreted stellar fraction leads to steeper density profile slopes (Figure 10, Paper I). The dependence of $\gamma_{\mathrm{outer}}$ and outer $h_{4}$ with environment is consistent with the previous findings that both stellar and halo masses anti-correlate with the steepness of the total density profile (Figures 4, 15, and 17 in Paper I), and fits in smoothly to the low-redshift ($z<0.5$) minor-merger-driven evolution phase of the ETG density profiles as discussed in Paper II (Figures 4, 7, 8). Combining the effects of stellar assembly on outer $h_{4}$, we conclude that the outer kinematic structure of ETGs co-evolves with their total density profiles, which is a natural consequence of hierarchical assembly in cosmological structure formation. The broad agreement of our results with the MASSIVE Survey ETGs also highlights the power of IllustrisTNG to elucidate the underlying correlations in a realistic formation scenario, that could not otherwise be discerned from observations alone.

Apart from the many aspects of our findings that do reproduce and explain the observed ETG kinematic features, we highlight that there are important systematics involving the central stellar velocity dispersion of the IllustrisTNG galaxies. We observe more galaxies with positive $\gamma_{\mathrm{inner}}$ compared to MASSIVE ETGs (Fig.~\ref{fig:4}) indicating underestimation of the central stellar velocity dispersion, which is consistent with the tendency of over-predicting central dark matter fraction (producing halo contraction) in IllustrisTNG galaxies~\citep{2018MNRAS.481.1950L,2020MNRAS.491.5188W}. This is also related to the negative $h_{4}$--$\beta_{\mathrm{3D}}$ correlation within $0.5 R_{\mathrm{eff}}^{\mathrm{MGE}}$ as seen in Fig.~\ref{fig:6}. These limitations pave a way forward for better understanding the interplay between dark matter and baryons, as well as for future improvements of the simulation subgrid physics model (especially related to AGN feedback), which could be better constrained by refined kinematic features at both large and small galactic radii. 

Another interesting feature is that the presence of infalling substructures can significantly boost $h_{3}$ and $h_{4}$ along the line-of-sight at the outskirts of ETGs (Appendix~\ref{sec:App_B}). Although quantifying the impact of mergers and their remnants on the kinematic structure of ETGs is beyond the scope of this work, this finding ties in neatly to our finding of the ETG outer kinematic structure by the fraction of stars acquired through minor perturbations. With ultra-high resolution IFU spectroscopy from MUSE~\citep{2010SPIE.7735E..08B} on the Very Large Telescope (VLT) or KCWI~\citep{2018ApJ...864...93M} on the Keck Telescope, non-Gaussian kinematic features could potentially provide a novel approach for discovering faint satellites around ETG hosts. Combined with upcoming photometric campaigns such as the Vera C. Rubin Observatory (LSST, \citealt{2019ApJ...873..111I}), the Nancy Grace Roman Space Telescope (WFIRST, \citealt{2015arXiv150303757S}), and Euclid~\citep{2012SPIE.8442E..0ZA} that advance in survey depth and field of view, we will soon be able to establish a more profound understanding of the co-evolution between the kinematic structure, density profile, and even the build up of the stellar halo around ETGs in the broad context of hierarchical structure formation. 

\section*{Acknowledgements}

We thank Dandan Xu, Ethan O. Nadler, Philip Mansfield, and Shengdong Lu for helpful discussions and support during the preparation of this paper. We thank the anonymous referee for insightful comments that helped to improve the draft. YW acknowledges the past support from the Tsinghua Xuetang Talents Programme and current support from a Stanford-KIPAC Chabolla Fellowship. This work is partly supported by the National Key Basic Research and Development Programme of China (No. 2018YFA0404501 to SM), and by the National Science Foundation of China (Grant No. 11821303 and 11761131004 to SM). MV acknowledges support through an MIT RSC award, a Kavli Research Investment Fund, NASA ATP grant NNX17AG29G, and NSF grants AST-1814053 and AST-1814259. 
VS acknowledges support by the Deutsche Forschungsgemeinschaft through project SP 709/5-1. This research made use of computational resources at the MIT/Harvard computing facilities supported by FAS and MIT MKI; the authors are thankful for the support from the FAS Research Computing team. The flagship simulations of the IllustrisTNG project used in this work have been run on the HazelHen Cray XC40-system at the High Performance Computing Center Stuttgart as part of project GCS-ILLU of the Gauss Centre for Supercomputing (GCS). This research made extensive use of \href{https://arXiv.org}{arXiv.org} and NASA's Astrophysics Data System for bibliographic information. 

\section*{Data Availability}

The IllustrisTNG public data can be accessed at \url{https://www.tng-project.org/data/}. The data in this article that are not part of the IllustrisTNG data release will be shared upon reasonable request to the corresponding author. 




\bibliographystyle{mnras}
\bibliography{Dyn_env}



\appendix

\section{The effective radius}
\label{sec:App_A}

\begin{figure}
	\includegraphics[width=\columnwidth]{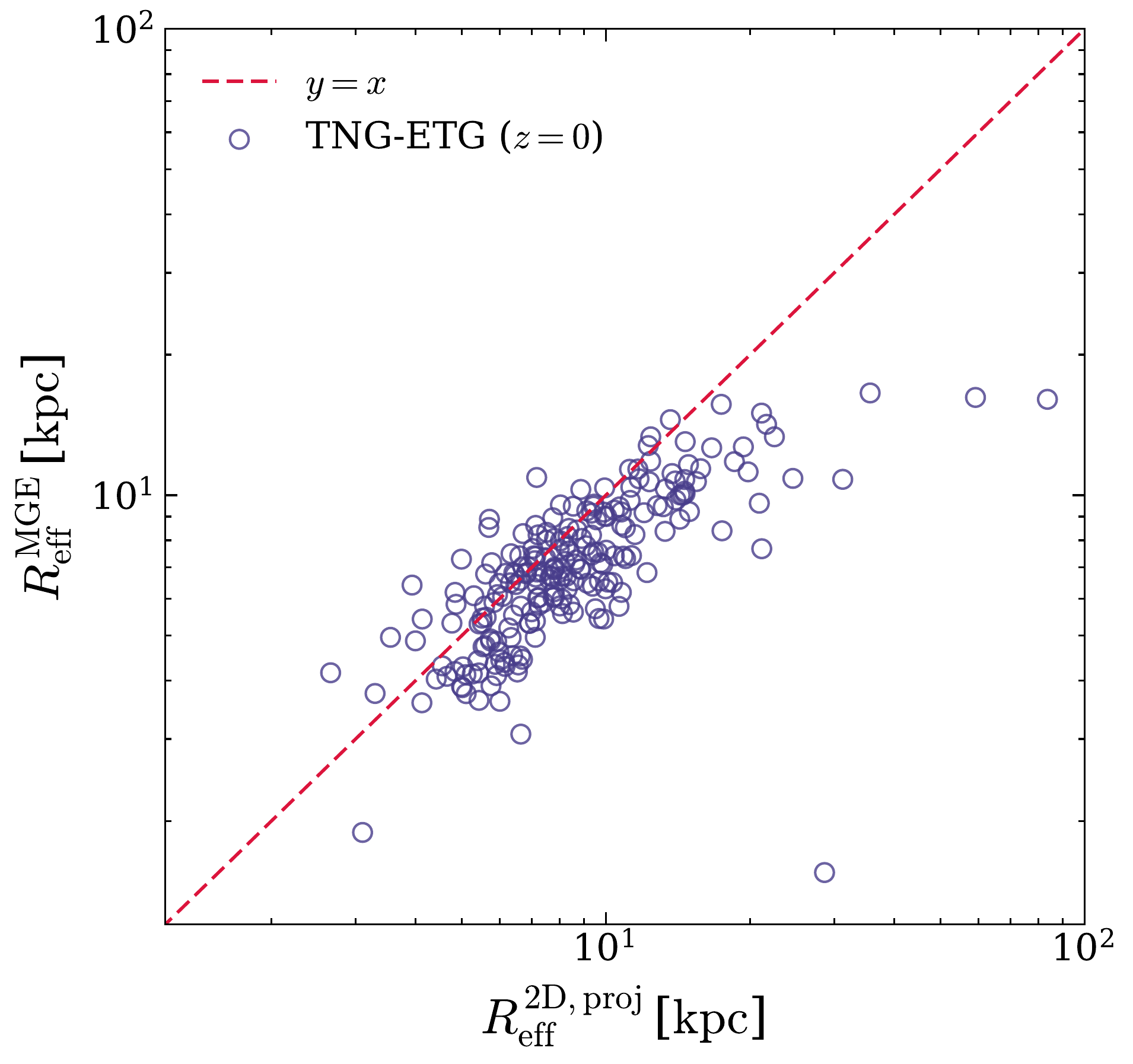}
    \caption{The comparison between the projected 2D effective radius calculated with all stellar particles assigned to the galaxy determined by \textsc{Subfind}, versus the MGE effective radius output by our mock observation pipeline. The MGE effective radii closely follow the 2D projected effective radii for galaxies with small sizes ($R_{\mathrm{eff}} \lesssim 10$ kpc). For the larger galaxies, the MGE effective radii are significantly smaller than the projected effective radii, which results from the MGE approach removing most of the diffuse intra-cluster light component assigned to the galaxy by \textsc{Subfind}, giving more realistic estimates of the effective radii for larger galaxies.}
    \label{fig:A1}
\end{figure}

\begin{figure*}
	\includegraphics[width=2\columnwidth]{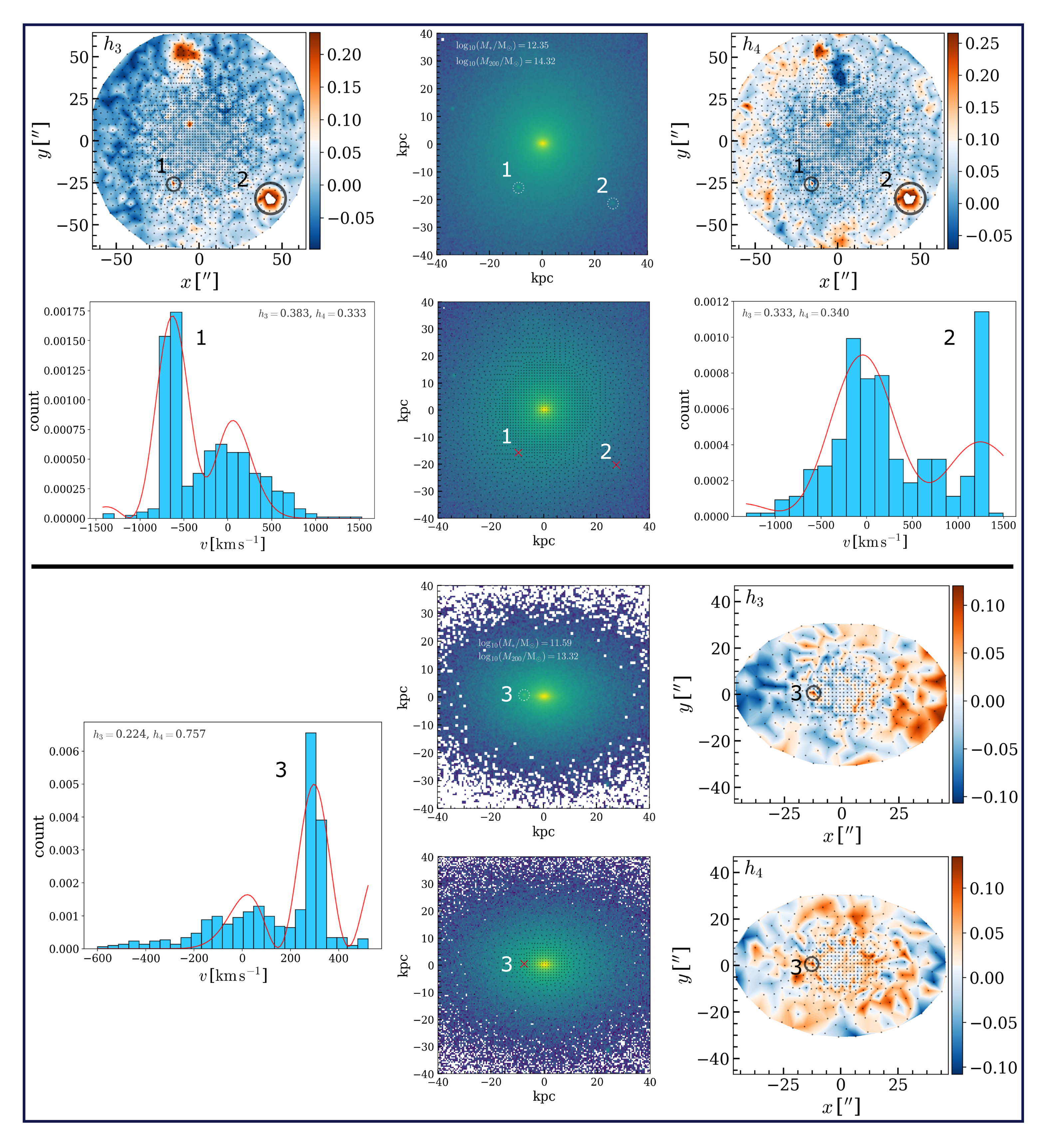}
    \caption{Three examples of spaxels/Voronoi bins cutting through infalling satellite galaxies around two different host galaxies that introduces large $h_{3}$ and $h_{4}$. {\it Upper panel:} Host \textsc{Subfind} ID 69507, with stellar mass $10^{12.35}\mathrm{M_{\astrosun}}$ and halo mass $10^{14.32}\mathrm{M_{\astrosun}}$. {\it Lower panel:} Host \textsc{Subfind} ID 257302, with stellar mass $10^{11.59}\mathrm{M_{\astrosun}}$ and halo mass $10^{13.32}\mathrm{M_{\astrosun}}$. The dashed circles in the luminosity maps indicate the location of these satellites, while the red crosses indicate the location of the spaxels/Voronoi bins covering them. Blue histograms show the LOS stellar velocity distribution in those bins (centered on the mean velocity, $v=0$, large peak offsets from the zero point indicate satellite stars), whereas the red curves indicate the $4^{\mathrm{th}}$ order Gauss-Hermite function fit. Their corresponding locations in the $h_{3}$ and $h_{4}$ maps are also marked by black circles. }
    \label{fig:C1}
\end{figure*}

In this section, we show the differences between the projected 2D effective radius ($R_{\mathrm{eff}}^{\mathrm{2D, proj}}$) and $R_{\mathrm{eff}}^{\mathrm{MGE}}$ of galaxies given by the best MGE fit. $R_{\mathrm{eff}}^{\mathrm{2D, proj}}$ is measured by projecting galaxies in random directions (along the z axis of the simulation box) and searching for the radius at which the enclosed projected stellar mass is half of the total stellar mass. However, this procedure includes all stellar particles assigned to galaxies by \textsc{Subfind}, and includes large contributions from the intra-cluster light (ICL) component around the massive ETGs that we selected. Another approach is to first model the galaxy's projected luminosity distribution in its central region (square aperture with 80 kpc side length) using Multi-Gaussian Expansion~\citep{1994A&A...285..739E}, and calculate the half-light radius using the set of analytic best-fit 2D Gaussians which eventually removes most of the ICL. It also has the advantage over simpler effective radius definitions in 2MASS (semi major axis length of an ellipse enclosing half the total galaxy luminosity) and the NSA catalog (half light radius from a single S$\mathrm{\acute{e}}$rsic fit) to flexibly preserve information of the spatial distribution for different components in the galaxy luminosity profile.

The comparison between the effective radii derived using these two methods is shown in Fig.~\ref{fig:A1}. For smaller galaxies ($R_{\mathrm{eff}} \lesssim 10$ kpc), the two definitions yield consistent size measurements, while for larger galaxies the MGE effective radii are significantly smaller than the 2D projected effective radii, which is due to the removal of the ICL. We adopt $R_{\mathrm{eff}}^{\mathrm{MGE}}$ in our comparison with observational data.

\section{Examples of minor mergers impacting the outer kinematic structure}
\label{sec:App_B}

In this section we showcase three examples of ongoing minor mergers introducing large non-Gaussian moments to the outer kinematic structure of ETGs. In Fig.~\ref{fig:C1}, we mark out three infalling satellites (overdensities in the luminosity maps) around two different host ETGs, with spaxels/Voronoi bins cutting through them along the line-of-sight, and resulting in significant $h_{3}$ and $h_{4}$ values. Clearly, these three pixels all demonstrate large peak offsets from the mean velocity (centered on $v=0$). While $h_{3}$ produced in this manner may diminish over time as the satellite orbits around the host due to its odd parity, a positive $h_{4}$ can be preserved over time due to its even parity. The cumulative effect of such in many minor mergers (along with flyby interactions that tidally heat galaxies, accelerating stars into the tails of the LOS velocity distribution) can gradually build up a global positive $h_{4}$ in the outskirts ($\gtrsim 1.5 R_{\mathrm{eff}}^{\mathrm{MGE}}$) of ETGs, leading to the positive correlation of the outer $h_{4}$ and stellar mass fraction from minor perturbations (minor mergers plus flyby interactions) as seen in Fig.~\ref{fig:7}.


\bsp	
\label{lastpage}
\end{document}